\documentclass{jkas}
\usepackage{siunitx}
\usepackage{amsmath}
\usepackage{amssymb}
\usepackage{physics}
\usepackage{hyperref}
\usepackage{subcaption}
\usepackage{caption}
\usepackage[dvipsnames]{xcolor}

\graphicspath{{./}{figures/}}

\def\year{2018} 
\def\month{April} 
\def\volume{00} 
\def\issue{0} 
\def\beginpage{1} 
\def\endpage{99} 
\def\received{XXX} 
\def\accepted{YYY} 
\date{Received \received; accepted \accepted}

\title{
Star Formation Activity of Galaxies Undergoing Ram Pressure Stripping in the Virgo Cluster
}

\author[1]{Jae Yeon~Mun}
\author[2]{Ho Seong~Hwang}
\author[1]{Myung Gyoon~Lee}
\author[3]{Aeree~Chung}
\author[3,4,5]{Hyein~Yoon}
\author[2]{Jong Chul~Lee}

\affil[1]{Department of Physics and Astronomy, Seoul National University, Gwanak-gu, Seoul 151-742, Korea; \email{jymarcie94@astro.snu.ac.kr}}
\affil[2]{Korea Astronomy and Space Science Institute, Daejeon 305-348, Korea; \email{hhwang@kasi.re.kr}}
\affil[3]{Department of Astronomy, Yonsei University, 50 Yonsei-ro, Seodaemun-gu, Seoul 03722, Korea}
\affil[4]{Sydney Institute for Astronomy, School of Physics, A28, The University of Sydney, Sydney, NSW 2006, Australia}
\affil[5]{ARC Centre of Excellence for All Sky Astrophysics in 3 Dimensions (ASTRO 3D)}

\def\corrauthor{
H. S. Hwang
}

\def\runningauthor{
Mun et al. 
}

\def\runningtitle{
SF Activity of Galaxies Undergoing RPS in Virgo
}

\def\keywords{
galaxies: clusters: general $-$ galaxies: clusters: individual (Virgo) $-$ galaxies: evolution $-$ galaxies: star formation
}

\def\abstracttext{
We study galaxies undergoing ram pressure stripping in the Virgo cluster to examine whether we can identify any discernible trend in their star formation activity. We first use 48 galaxies undergoing different stages of stripping based on HI morphology, HI deficiency, and relative extent to the stellar disk, from the VIVA survey. We then employ a new scheme for galaxy classification which combines HI mass fractions and locations in projected phase space, resulting in a new sample of 365 galaxies. We utilize a variety of star formation tracers, which include \textit{g $-$ r}, \textit{WISE} [3.4] $-$ [12] colors, and starburstiness that are defined by stellar mass and star formation rates to compare the star formation activity of galaxies at different stripping stages. We find no clear evidence for enhancement in the integrated star formation activity of galaxies undergoing early to active stripping. We are instead able to capture the overall quenching of star formation activity with increasing degree of ram pressure stripping, in agreement with previous studies. Our results suggest that if there is any ram pressure stripping induced enhancement, it is at best locally modest, and galaxies undergoing enhancement make up a small fraction of the total sample. Our results also indicate that it is possible to trace galaxies at different stages of stripping with the combination of HI gas content and location in projected phase space, which can be extended to other galaxy clusters that lack high-resolution HI imaging. 
}

\begin{document}
\jkashead 


\section{Introduction\label{sec:intro}}
Star formation and nuclear activity of galaxies are strongly affected by the cluster environment \citep{vonderlinden10, hwang12a} through various physical mechanisms: ram pressure stripping \citep{gunngott72}, thermal evaporation \citep{cosong77}, strangulation \citep{larson80, bekki02}, galaxy harassment \citep{moore96}, and cumulative galaxy-galaxy hydrodynamic/gravitational interactions \citep{parkhwang09}. Among them, ram pressure stripping is one of the most effective ways to remove gas from galaxies. Ram pressure arises due to the hot X-ray gas embedded in the intracluster medium (ICM), and pushes against the interstellar medium (ISM) of a galaxy as the galaxy travels through the galaxy cluster. As a galaxy approaches the cluster core, the strength of ram pressure increases, and a greater amount of its gas - the fuel necessary for star formation - is stripped \citep{gunngott72}. In general, ram pressure stripping (RPS) is expected to quench star formation in cluster galaxies (e.g., \citealt{kk04b, boselli06, crowl08, jaffe15}). However, there have been observational and theoretical studies insinuating the possibility of ram pressure induced star formation enhancement prior to quenching (e.g., \citealt{botdres86, fujita99, kronberger08, kapferer09}). One-sided tails of H$\alpha$ emission or young stellar knots extending beyond disks of galaxies in distant clusters have shown to be linked to interactions with the ICM (e.g., \citealt{cortese07, hester10, smith10}), giving further support to the scenario of ram pressure induced star formation enhancement. While not all galaxies subject to ram pressure stripping show such features, those that exhibit extended tails of optically bright gas are known as ``jellyfish" galaxies. As of recent, many observational studies of jellyfish galaxies have captured the enhancement of star formation activity in their disks and tails \citep{vulcani18, ramatsoku19}. In particular, \citet{vulcani18} categorized their sample of jellyfish galaxies from the GAs Stripping Phenomena survey (GASP; \citealt{poggianti17b}) into different stripping stages - initial, moderate, and extreme stripping, along with truncated disks - based on their H$\alpha$ morphology. Upon comparing the star formation activity of their sample to field galaxies of similar stellar mass, they show that jellyfish galaxies undergoing moderate to extreme stripping show the strongest enhancement of star formation. As such, it is of great interest to probe the change in star formation activity of a galaxy as it undergoes different stages of ram pressure stripping.

Given its close proximity, the Virgo cluster has often been the target of observational studies of interstellar medium (ISM) - intracluster medium (ICM) interactions (e.g., \citealt{boselli18}). \citet{kk04b} studied the H$\alpha$ morphology of Virgo cluster and isolated spiral galaxies to not only divide cluster galaxies into different categories - normal, anemic, enhanced, and truncated - but also to identify the history of environmental processes that galaxies may have experienced in the cluster environment. While H$\alpha$ is a useful probe of the current star formation activity, the distribution of neutral hydrogen (HI) of a galaxy offers a clearer picture of the environmental processes a galaxy has undergone. Given that HI gas is usually extended out towards the outer disk region of a galaxy, it is often the first component to show perturbations due to external processes such as tidal interactions and ram pressure stripping. \citet{chung09} obtained high-resolution HI data of a sample of 53 late type galaxies in the Virgo cluster, among which 46 of them were selected based on the classification in \citet{kk04b} to probe galaxies with a wide range of star formation properties. Known as the VLA Imaging survey of Virgo galaxies in Atomic gas (VIVA), \citet{chung09} identified a trend in HI morphology with respect to the projected location of galaxies within the cluster. Galaxies near or in the core were often found with symmetric HI disks truncated with respect to their stellar disks, while galaxies in the outskirts were generally HI-rich in comparison (but still HI poor compared to field galaxies). On the other hand, galaxies located at intermediate distances from the cluster core (i.e., M87; 0.6 $\lesssim$ $d_{M87}$ $\lesssim$ 1 Mpc) were observed to have one-sided HI tails pointing away from the core, which led \citet{chung07} to suggest that such galaxies have only recently began falling into the cluster, and that the impact of ram pressure on the gas content of galaxies extends out towards intermediate distances.

It is then possible to infer from the results of previous studies that the combination of HI morphology and projected location from the cluster center can offer a closer look into the orbital history of a galaxy within a galaxy cluster. As of recent, many studies have incorporated the use of projected clustercentric distance and line-of-sight velocity with respect to the cluster mean in the form of a phase-space diagram to identify galaxies at different stages in their orbits \citep{oman13, oman16, rhee17}. For example, \citet{jaffe15} studied Abell 963, a galaxy cluster at z = 0.2, by combining the availability of HI detections and location in projected phase space. They found that HI-detected galaxies were generally located in the ``recent infalls'' region, whereas there was a noticeable lack of HI-detections within the ``stripped'' and ``virialized'' regions. Moreover, they investigated the stellar populations of their galaxy sample using \textit{NUV $-$ r} color to find that the bluest galaxies were generally found outside the ``stripped'' and ``virialized'' zones, while the redder galaxies seemed to dominate those two regions. While they were not able to take HI morphology into account for their analysis, their results confirm the general trend that as blue, star-forming galaxies fall into a cluster, its gas content will be gradually stripped by ram pressure as it settles, with its star formation being quenched in the process. 

Given that observational studies of jellyfish galaxies have shown that a stage of star formation enhancement could occur prior to its gas content being completely quenched, this then brings up the question of whether a similar analysis can be done for the Virgo cluster. \citet{yoon17} provides an ideal sample to answer such a question, as they have used the VIVA \citep{chung09} sample to categorize galaxies in the Virgo cluster into different stages of stripping based on their HI morphology, HI deficiency (def\textsubscript{HI}, a measure of how deficient in HI a galaxy is relative to its field counterpart of the same size, \citealt{haynes84}), and relative extent to the optical disk. The availability of high-resolution HI imaging plays a crucial role in their study, as the peculiarities in the HI morphology allows a direct probe of the environmental processes that the galaxy may have been subject to. \citet{yoon17} also briefly examined the star formation properties of the categorized galaxies using the H$\alpha$ equivalent width and H$\alpha$ concentration from \citet{kk04a}. Their results \citep[see][Fig. 8]{yoon17} showed reduced star formation and increased concentration of star formation in the galaxy center, along with decreasing HI gas content from early through post stripping stages. 

In this work, we aim to build upon the work of \citet{yoon17} by examining the star formation properties of their sample of galaxies in more detail, using various types of star formation tracers. To make up for the small sample size in their study, we expand the sample by using the Extended Virgo Cluster Catalog (EVCC; \citealt{kim14}). Since high-resolution HI data are not available for the entirety of the EVCC, we make use of HI mass fraction (relative HI mass to stellar mass, M\textsubscript{HI}/M\textsubscript{*}) and location in projected phase space. While  M\textsubscript{HI}/M\textsubscript{*} can work as a first-order indicator of gas stripping, the location of galaxies in phase space assures that the origin of stripping is likely to be of ram pressure \citep{rhee17, jaffe15, yoon17}. 

This paper is organized as follows. In Section \ref{sec:data}, we describe the sample and explain the derivation of physical properties used to examine the star formation activity of galaxies in the Virgo cluster. In Section \ref{sec:results}, we explain how the different stripping classes were defined in both \citet{yoon17} and this study, following up with the analysis of the star formation activity per class. In Section \ref{sec:disc}, we discuss the limitations of integrated properties in capturing local star formation enhancements in Virgo galaxies, and how the simplification of the classification scheme using HI gas fraction and location in phase space allowed the trend of quenching of star formation activity to be observed with a larger sample. We summarize our results in Section \ref{sec:con}. Throughout this paper, we adopt a Salpeter \citep{salpeter55} initial mass function (IMF) and assume a $\Lambda$ cold dark matter cosmology with $\Omega_{M}$ = 0.3, $\Omega_{\Lambda}$ = 0.7, and H\textsubscript{0} = 70 km s\textsuperscript{-1} Mpc\textsuperscript{-1}. We also adopt a distance of 16.5 Mpc, or ($m - M$) = 31.1, to the Virgo cluster \citep{mei07}. All photometric magnitudes used are given in the AB magnitude system.

\section{Data} \label{sec:data}
\subsection{EVCC} \label{subsec:galcat}
The EVCC is comprised of fundamental information such as cluster membership, morphology, radial velocities, and $ugriz$ photometry of a total of 1589 galaxies spread throughout the main body to the outskirts of the Virgo cluster. More precisely, the spatial coverage of the EVCC extends out to $\sim$4 times the virial radius ($r_{200}$ = 1.55 Mpc, \citealt{mclaugh99, ferrarese12}) of the cluster, for which \citet{kim14} excluded background galaxies with radial velocities larger than 3000 km s\textsuperscript{-1}.

With regards to the $ugriz$ photometry, the Sloan Digital Sky Survey (SDSS) photometric pipeline is known to be unreliable for large and bright galaxies due to issues with deblending and sky subtraction \citep{sdssdr2, sdssdr7, west10}. To account for such shortcomings, \citet{kim14} performed their own Source Extractor (SExtractor, \citealt{sextrac}) photometry on $ugriz$ images from SDSS Data Release 7 (DR7). As shown by Figure 13 in \citet{kim14}, the growing discrepancy between the  photometric magnitudes with increasing angular size of a galaxy further confirms that the total flux derived from the SDSS pipeline does not give an accurate representation of the total light from nearby galaxies such as those in the Virgo cluster. We thus attest that physical parameters derived using EVCC \textit{ugriz} photometry, such as stellar mass, are reliable measurements of the stellar population of the galaxies in the Virgo cluster. 

\subsection{Physical Parameters of Galaxies} \label{subsec:physpar}
To examine the star formation properties of Virgo cluster galaxies, we proceed by building a master catalog based on the EVCC. We add stellar masses, star formation rates (SFRs), mid-infrared photometric data, along with neutral hydrogen masses to the catalog. Here, we briefly describe each of the physical parameters.

\subsubsection{Stellar Masses} \label{subsubsec:mstar}
The stellar masses of EVCC galaxies are computed using \textit{Le Phare} \citep{arnouts99, ilbert06}. We used EVCC $ugriz$ photometry and redshifts to build synthetic spectral energy distributions (SEDs) from the stellar population synthesis (SPS) models of Bruzual $\&$ Charlot (BC03; \citeyear{bc03}), which are based on the Chabrier \citeyearpar{chabrier03} IMF. The synthetic SEDs are computed at different redshifts with $\delta$\textit{z} = 0.001 as the furthest galaxy in our sample is located at z $\sim$ 0.01. The BC03 templates provided by \textit{Le Phare} have three metallicities and seven exponentially decreasing star formation models (SFR $\propto$ $e^{-t/\tau}$) with $\tau$ = 0.1, 0.3, 1, 2, 3, 5, 10, 15, and 30 Gyr. We allow the age of the stellar population to vary between 0 and 13 Gyr. Dust extinction was applied to the SPS templates using the extinction law of \citet{calzetti00}, with values of $E(B - V)$ varying from 0 to 0.6. Adopting different values of the stellar population parameters gives rise to a distribution of different estimates for the stellar mass of each galaxy. We adopt the median of the distribution as the stellar mass of each galaxy. 

\begin{figure*}[htb!]
\centering
\includegraphics[width=\textwidth]{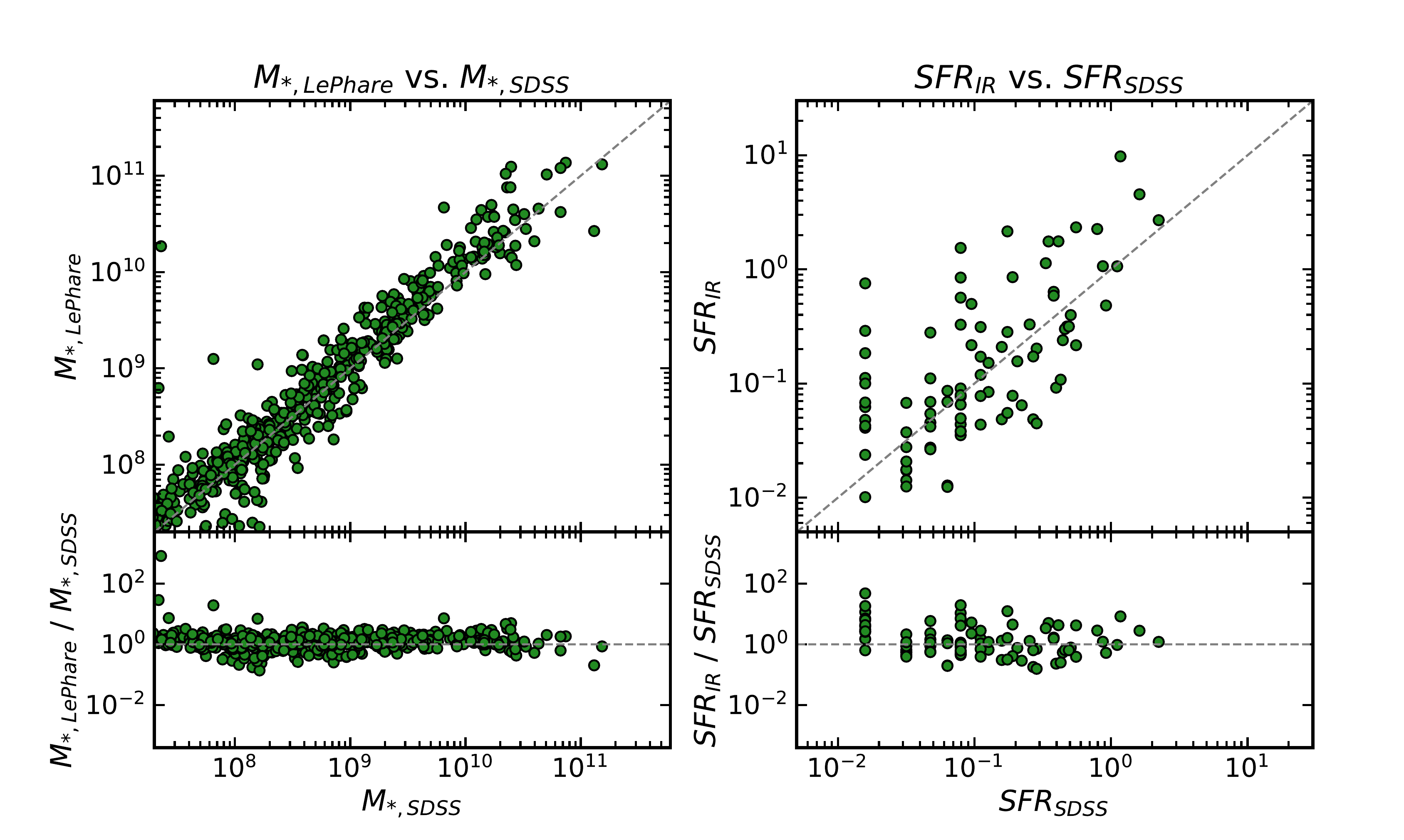}
\caption{Comparison diagram of stellar masses measured via an SED fit of EVCC \textit{ugriz} photometry with \textit{Le Phare} (left), and star formation rates derived using total IR luminosities with those given in the SDSS MPA/JHU DR7 VAGC. The dashed lines in the top panels show the one-to-one relations. \label{fig:fig1}}
\end{figure*}

We compare our stellar mass estimates with those given by the SDSS MPA/JHU DR7 VAGC \citep{kauff03}, where stellar masses were measured based on a SED fit of SDSS \textit{ugriz} photometry with the SPS models of \citet{bc03}. The stellar mass estimates from the MPA/JHU DR7 VAGC are based on the Kroupa IMF. For consistency throughout this study, stellar mass estimates computed via \textit{Le Phare} and those given by the MPA/JHU DR7 VAGC are converted to a Salpeter \citeyearpar{salpeter55} IMF by multiplying the former by a factor of 10\textsuperscript{0.25} \citep{chabrier03} and the latter by a factor of 10\textsuperscript{0.2} \citep{kroupa01}. Both conversion factors were taken from \citet{bernardi10}. We show the results of the comparison in the left panel of Figure \ref{fig:fig1}, which shows good agreement between the two stellar mass estimates with a median offset of about 0.08 dex. While there is in general a good agreement between the two methods, the rms of the difference between our estimates and those given by the MPA/JHU DR7 VAGC is about 0.23 dex when including all the galaxies in common. The absolute uncertainty arising from panchromatic SEDs is approximately 0.3 dex \citep{conroy13}, which suggests that our stellar mass estimates are overall consistent with the measurements given by the SDSS MPA/JHU DR7 VAGC.

\subsubsection{Star Formation Rates \& Mid-Infrared Colors} \label{subsubsec:sfrwise}
We use mid-infrared photometric data from the \textit{Wide-field Infrared Survey Explorer} (\textit{WISE}, \citealt{wright10}) to derive SFRs and examine mid-infrared colors. The \textit{WISE} survey covers the entire sky in the mid-infrared, offering four-band photometric data (3.4, 4.6, 12, and 22$\mu$m) for over 747 million objects. We used a matching tolerance of \SI{6}{\arcsecond} ($\sim$FWHM of the PSF at 22$\mu$m) between the EVCC and AllWISE Source Catalog\footnote{\url{http://wise2.ipac.caltech.edu/docs/release/allwise/}}. This matching tolerance is small enough to ensure that we avoid matching our sample with the wrong mid-infrared counterparts. In the cross-matching process, we found some issues with the coordinates given by the EVCC. EVCC provides two different coordinates, one derived from SExtractor, another given by the SDSS fiber locations. The coordinates given by SExtractor were offset from the center of nearby galaxies, resulting in an offset as large as $\sim$\SI{13.68}{\arcsecond} in one particular case. We found that coordinate offsets occur for large spiral galaxies given the presence of many clumps in their disks, despite the fact that the photometry given by EVCC was carefully conducted. Given that the galaxies with the largest offsets were those with high-resolution HI imaging used in \citet{yoon17}, it was necessary to ensure that these galaxies could be found within a matching tolerance of \SI{6}{\arcsecond}. We thus replaced the coordinates of \citeauthor{yoon17}'s 48 galaxies with those given by the NASA Extragalactic Database (NED). For 2 galaxies with no match, the NED-given coordinates are replaced with \textit{WISE}-given coordinates of the closest \textit{WISE} detection in the AllWISE Source Catalog. There were 6 additional galaxies in the EVCC that were not found within the matching tolerance, which were manually searched for and added to the master catalog. The AllWISE Source Catalog provides both point source profile-fitting and elliptical aperture magnitudes. Because we are dealing with nearby galaxies, some of them are well-resolved to the point that we generally require larger apertures to obtain an accurate measurement of their total brightness. We thus prioritize the use of elliptical aperture magnitudes for galaxies in the EVCC. For those with no elliptical magnitudes given by the AllWISE Source Catalog (i.e., relatively small galaxies), we use point source profile-fitting magnitudes instead. We also select reliable flux density values with a signal-to-noise ratio cut of S/N $\geq$ 3 in each band. 

\begin{table*}[t]
\caption{The VIVA sample defined by different stages of gas stripping \label{tab:tab1}}
\captionsetup{justification=justified}
\centering
\begin{tabular}{cccl}
\toprule
    Class & def\textsubscript{HI} & def\textsubscript{HI} & HI morphology and deficiency compared to field galaxies \\
     & (Range) & (Median) & \\
    (1) & (2) & (3) & (4) \\ 
\midrule
    Class 0 & -0.81 $\sim$ 0.38 & 0.13 & No definite signs of gas stripping due to the ICM \\
    (N = 13) & & & \\
    & & & \\
    Class I & -0.43 $\sim$ 0.41 & 0.02 & One-sided HI feature; \\
    (N = 7) & & & relatively comparable to field galaxies in HI gas content \\
    & & & \\
    Class II & 0.12 $\sim$ 1.16 & 0.76 & Highly asymmetric HI disk; quite deficient in HI, \\
    (N = 10) & & & with an average of $\sim$17$\%$ of HI mass \\
    & & & compared to those of field galaxies.\\
    & & & \\
    Class III & 0.82 $\sim$ 2.25 & 1.42 & Symmetric and severely truncated HI disk; \\
    (N = 10) & & & extremely deficient in HI, \\
    & & & with $<$4$\%$ of the HI mass of a field counterpart. \\
    & & & \\
    Class IV & 0.51 $\sim$ 1.17 & 0.79 & Symmetric HI disk with marginal truncation \\ 
    (N = 8) & & & within the radius of the stellar disk;  \\
    & & & lower HI surface density than other classes; \\
    & & & quite deficient in HI with on average $\sim$15$\%$ \\
    & & & of the HI mass of a field counterpart. \\
\bottomrule
\end{tabular}
\caption*{(1) RPS class, (2) HI deficiency range, (3) median HI deficiency, (4) HI morphology and relative HI deficiency compared to those of field galaxies. Table adapted from \citet{yoon17}.}
\end{table*}

To compute SFRs, we use \textit{WISE} 22$\mu$m flux densities, which were shown to be reliable SFR indicators in \citet{hwang12c} and \citet{lee13}. We derive total infrared (IR) luminosities from \textit{WISE} 22$\mu$m luminosities using the SED templates of Chary $\&$ Elbaz (\citeyear{ce01}). We then convert the total IR luminosities into SFRs with the Kennicutt \citeyearpar{kennicutt98} relation, assuming a Salpeter IMF: SFR\textsubscript{IR} (M\textsubscript{$\odot$} yr\textsuperscript{-1}) = 1.72 $\times$ 10\textsuperscript{-10} L\textsubscript{IR} (L\textsubscript{$\odot$}). We remove active galactic nuclei (AGN) from the sample to avoid contamination in the mid-infrared. We identify 21 AGN via the Baldwin-Phillips-Terlevich (BPT) classification provided by the MPA/JHU DR7 VAGC. We also use the \textit{WISE} mid-infrared color-color selection criteria defined by \citet{jarrett11} and \citet{mateos12} to identify one additional AGN. A total of 24 AGN, including 2 composite galaxies with large offsets from the one-to-one relation in Figure \ref{fig:fig1} (via 2-sigma clipping), were identified and removed from the sample.

We then compare our SFR estimates with those in the MPA/JHU DR7 VAGC \citep{brinch04} in the right panel of Figure \ref{fig:fig1}. The SFRs in the MPA/JHU DR7 VAGC were derived from extinction and aperture-corrected H$\alpha$ luminosities. For AGN and galaxies with weak emission lines, they measure the SFRs using the \SI{4000}{\angstrom} ($D_{n}$4000) break. There seems to be a correlation between the two measurements despite a large scatter. We caution the reader that the two measurements are not guaranteed to show a tight correlation, as it is not ideal to derive physical parameters for galaxies in our sample’s redshift range, due to the small size of the SDSS fibers. \citep{kewley05}. As such, the plot should be referenced simply for comparison purposes. The median offset between SFR\textsubscript{IR} and SFR\textsubscript{SDSS} is approximately 0.03 dex. The data with the largest offsets can primarily be explained by two factors. Upon visual inspection of SDSS images and examination of optical size estimates, we were able to conclude that most galaxies with measured SFR\textsubscript{IR} larger than SFR\textsubscript{SDSS} were either optically large or contained inner structures (e.g., bar, bulge). We attribute the likely cause of this discrepancy to the fact that SDSS fiber observations were limited to the central part of the galaxies with low star formation activity, thus leading to an underestimation of the SFR despite aperture corrections. On the other hand, we found that most galaxies with SFR\textsubscript{IR} smaller than SFR\textsubscript{SDSS} were blue irregular galaxies. This makes sense as UV bright galaxies that consist entirely of young stellar populations (and thus do not contain much dust) would result in a very weak detection in \textit{WISE} 22$\mu$m. This would in turn result in a higher measurement of the SFR in H$\alpha$ and a lower measurement of the SFR in the infrared.

\subsubsection{Atomic Gas Masses} \label{subsubsec:gas}
We adopt the atomic gas masses, M\textsubscript{HI}, from the Arecibo Legacy Fast ALFA Survey (ALFALFA, \citealt{giovanelli05a, haynes11, haynes18}). The latest release, the ALFALFA Extragalactic HI Source catalog \citep{haynes18}, provides HI data for a total of $\sim$31,500 extragalactic sources detected out to z $<$ 0.06. We cross-match the catalog with the EVCC within a matching tolerance of \SI{4}{\arcsecond}, giving a total of 579 matched objects. Among the 579 matched objects, there are a total of 39 galaxies with RPS classes defined by \citet{yoon17}. For the rest of the RPS class sample without HI masses from ALFALFA, we adopt HI masses from the VIVA survey. The HI detection limit for the ALFALFA survey is about 10\textsuperscript{6} M\textsubscript{$\odot$}, and the stellar mass of our sample ranges from about 2$\times$10\textsuperscript{7} M\textsubscript{$\odot$} to 6$\times$10\textsuperscript{11} M\textsubscript{$\odot$}. This indicates that the lowest valid value of the relative HI-to-stellar mass ratio for our sample is approximately 2$\times$10\textsuperscript{-6}, which is equivalent to a logarithmic value of -5.8. Our lowest measured value gives a logarithmic value of -2.97, suggesting that all our measurements of the HI mass fractions are valid.

\begin{figure*}[htb!]
\centering
    \begin{subfigure}[b]{0.48\textwidth}
        \includegraphics[width=\textwidth]{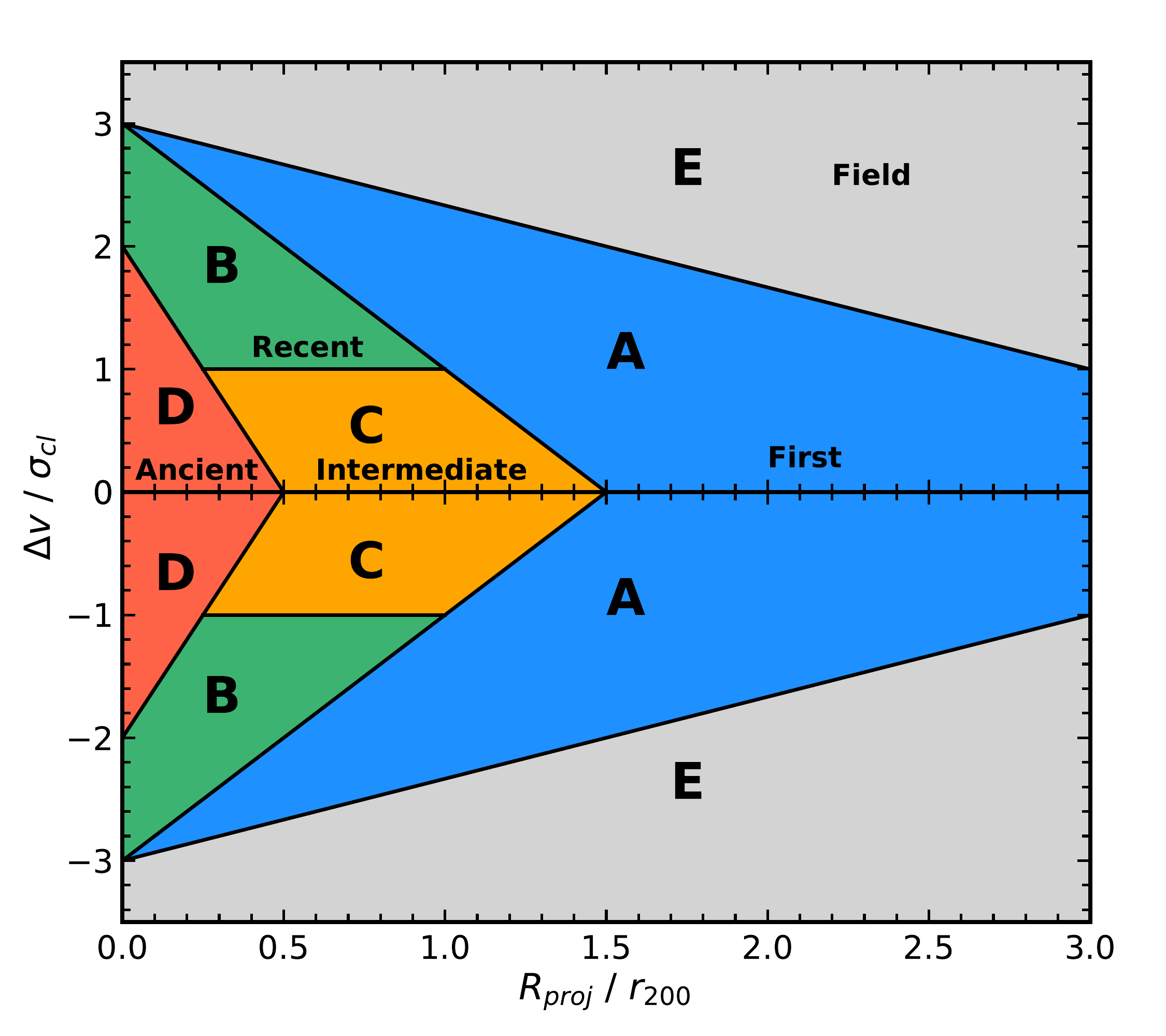}
    \end{subfigure}
    \begin{subfigure}[b]{0.48\textwidth}
        \includegraphics[width=\textwidth]{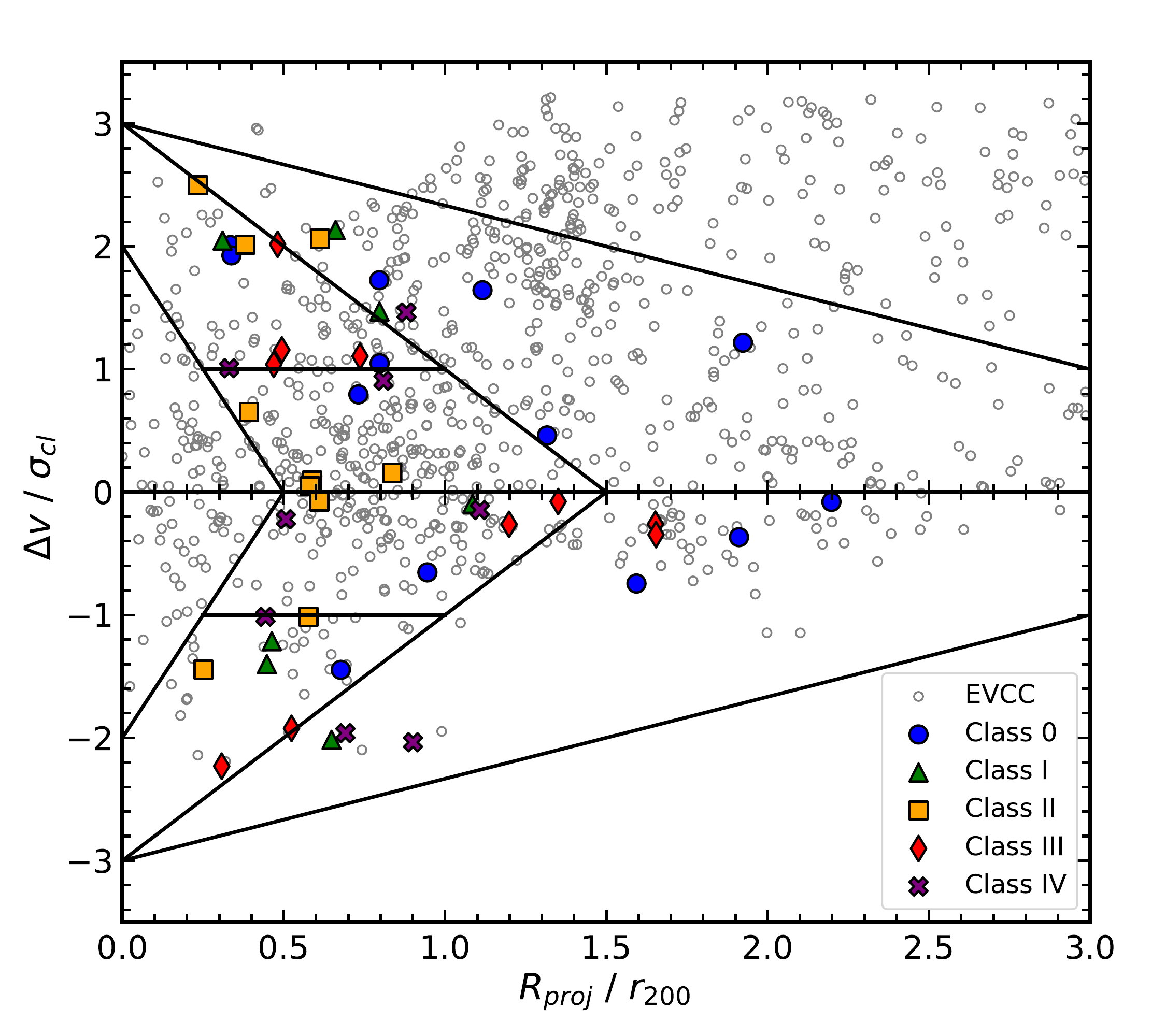}
    \end{subfigure}
\caption{(Left) Projected phase-space diagram with regions A $\sim$ E defined according to time since infall into the cluster center, closely following the classification in \citet{rhee17}. Clustercentric distance and velocity are normalized by the cluster virial radius $r_{200}$ and cluster velocity dispersion $\sigma_{cl}$, respectively. First infallers are located in Region A, recent infallers in Region B, intermediate infallers in Region C, and ancient infallers in Region D. Region E is defined by field galaxies. (Right) Projected phase-space diagram of RPS class galaxies and the rest of the EVCC sample shown as reference (grey open symbols). \label{fig:fig2}}
\end{figure*}

\section{Results} \label{sec:results}
\subsection{Star Formation Properties of Galaxies Undergoing Different Stages of RPS} \label{subsec:rps}
\subsubsection{Definition of RPS Classes} \label{subsubsec:rpsdef}
\citet{yoon17} categorize a total of 35 galaxies into different stages of stripping and identify an additional class of 13 control galaxies that are not subject to stripping. Characterization of each class is briefly described in Table \ref{tab:tab1}, adapted from Table 1 and Section 2 in \citealt{yoon17}. Among the stripping classes, Classes I, II, and III represent galaxies undergoing early, active, and post stripping stages, respectively. On the other hand, galaxies that show no strong evidence of stripping due to the ICM are utilized as a control sample, labeled as Class 0 (N = 13), whereas galaxies that appear to have lost HI gas at all radii due to a mechanism other than ram pressure stripping, such as starvation \citep{larson80}, are labeled as Class IV (N = 8). 

\begin{figure*}[htb!]
\centering
\includegraphics[width=0.8\textwidth]{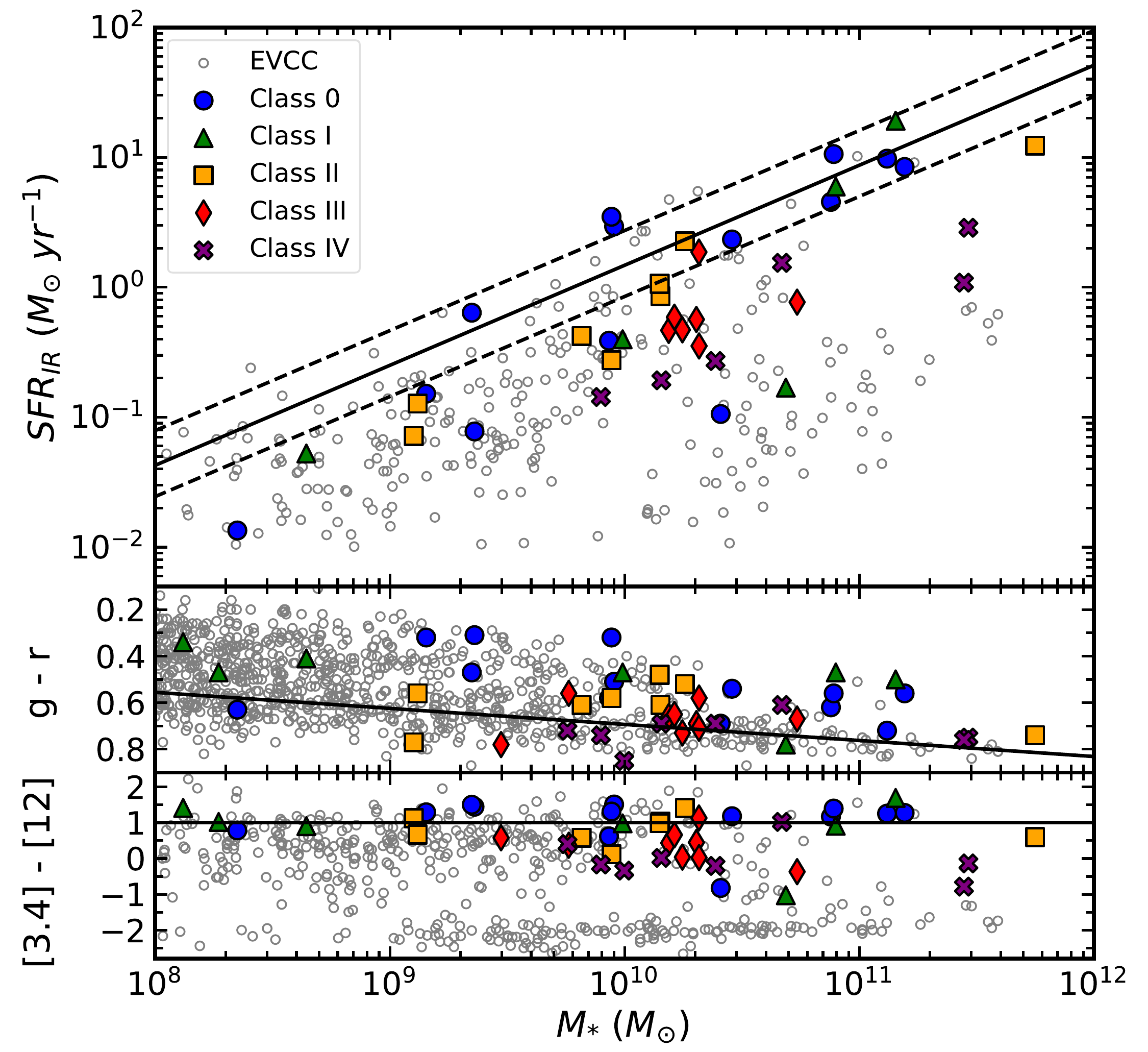}
\caption{Total IR-derived star formation rates, optical, and mid-infrared colors (from top to bottom) as a function of stellar mass for Virgo galaxies with defined RPS classes from \citet{yoon17} and the remaining EVCC galaxies with no defined classes plotted as reference. The solid black lines define a reference sequence for each star formation activity tracer, to normalize for the effects of different masses. The reference sequence for SFR\textsubscript{IR} is the star-forming main sequence defined by \citet{elbaz07}, and the dashed lines are the lower and upper bounds to the fit. The solid line in the middle panel is the red sequence, which was derived based on a best linear fit of non-RPS Class galaxies with M\textsubscript{*} $>$ 1.5$\times$10\textsuperscript{10} M\textsubscript{$\odot$} and \textit{g $-$ r} $>$ 0.58. The solid line in the bottom panel is adopted from the peak of the mid-infrared color distribution of star-forming galaxies in \citet{hwang12b}. \label{fig:fig3}}
\end{figure*}

\citeauthor{yoon17} utilize these five classes in combination with projected phase space to better trace the orbital histories of HI stripped galaxies within the Virgo cluster. Given the large scatter in projected phase space due to projection effects (e.g., \citealt{oman13, rhee17}), the availability of high-resolution HI imaging plays an important role in identifying the orbital stage of the selected galaxies. We plot the locations of RPS class galaxies, along with other EVCC galaxies, in projected phase space in the right panel of Figure \ref{fig:fig2}. In the diagram, we can clearly see that our reference sample is not uniformly distributed in projected phase space, such that they preferentially inhabit regions with positive clustercentric velocities. We attribute the cause of this non-uniformity to the limited sky coverage of the EVCC. \citeauthor{yoon17} find reference galaxies traveling at positive velocities to be consistent with the NGC 5353/4 filament and the W-M sheet \citep{kim16}, which are found to be located behind the Virgo cluster. On the other hand, \citet{kim16} also find filamentary structures located in front of and/or in the vicinity of Virgo, to be elongated towards the cluster, indicating cluster infall. However, these structures fall outside of the sky coverage of the EVCC, which explains the absence of galaxies in the bottom half of the phase-space diagram.

In the left panel of Figure \ref{fig:fig2}, we label different regions in projected phase space according to time since infall, using a modification of the classification adopted by \citet{rhee17}. According to the expected orbit of a radially infalling galaxy (e.g., \citealt{yoon17, rhee17, jaffe15}), a galaxy under the influence of the cluster potential will slowly accelerate from the outskirts to eventually virialize in the cluster core after multiple core crossings. Classes I $\sim$ III overall follow the expected trend of radially infalling galaxies within projected phase space, as shown in the right panel of Figure \ref{fig:fig2}. Class I galaxies are located at high velocity and low to intermediate clustercentric distances, which in combination with their HI morphology, show that they are likely to be on their first infall into the cluster core. Class II galaxies are at similar locations in projected phase space as Class I galaxies, with a select few found to be traveling at relatively low velocities. For the latter, \citeauthor{yoon17} suggest that tidal interactions may cause a shift in projected phase space, and thus caution is advised when interpreting their locations with respect to their orbital histories. Although some Class III galaxies are found in the same region in projected phase space as Class II/I objects, their heavily truncated HI disks suggest that they have already experienced their first core crossing, and are now falling out of the core. HI morphology alone suggests that Class III galaxies are at a later stripping stage than Class II galaxies are. However, their relative locations in projected phase space imply that a subsample of Class II galaxies are at a later stage in their orbital histories relative to some Class III galaxies. \citeauthor{yoon17} therefore suggest that Class III galaxies do not always occur after Class II. Among the Class III galaxies, there are a subsample of them found at radii beyond the cluster virial radius (i.e., $R_{proj}$/$r_{200}$ = 1). \citeauthor{yoon17} state that their low velocities with respect to the cluster, on top of them being located outside the cluster virial radius, suggest that they have fallen out of the cluster core after first infall, hence labeling them as a ``backsplashing'' population. While Class IV galaxies are generally found at low to intermediate clustercentric distances, they are rather spread out in projected phase space. Class 0 galaxies also exhibit a large scatter in projected phase space, although more than half of them are found in the first infalls (i.e., Region A) region. With the exception of galaxies subject to other stripping mechanisms such as tidal interactions and starvation, or those that do not seem to show any signs of gas depletion, \citeauthor{yoon17} demonstrate that the combination of HI morphology and location in projected phase space enriches the information one can obtain with regards to the orbital history of galaxies within a galaxy cluster.

\begin{figure*}[htb!]
\centering
    \begin{subfigure}[b]{0.3\textwidth}
        \includegraphics[width=\textwidth]{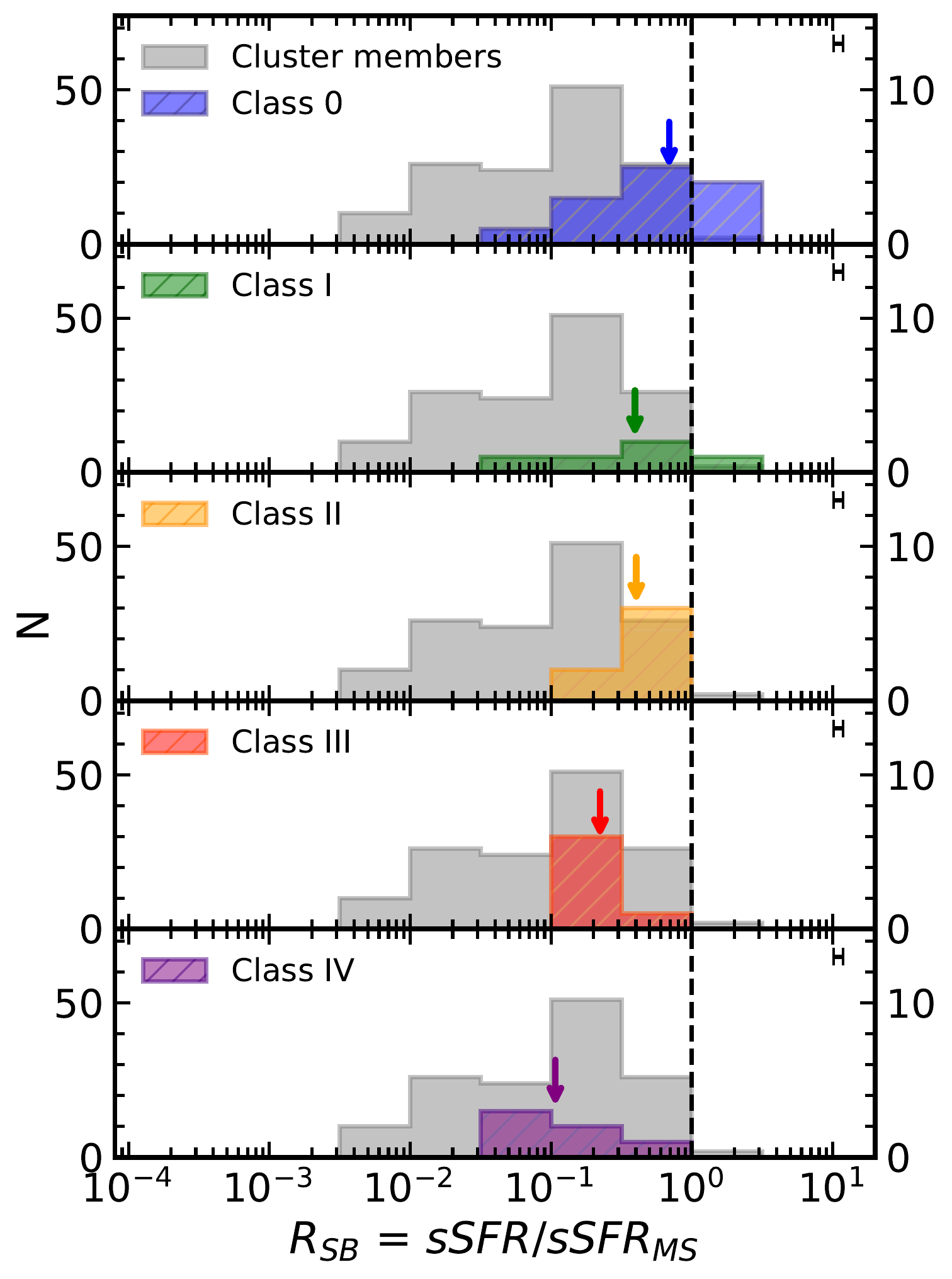}
    \end{subfigure}
    \begin{subfigure}[b]{0.3\textwidth}
        \includegraphics[width=\textwidth]{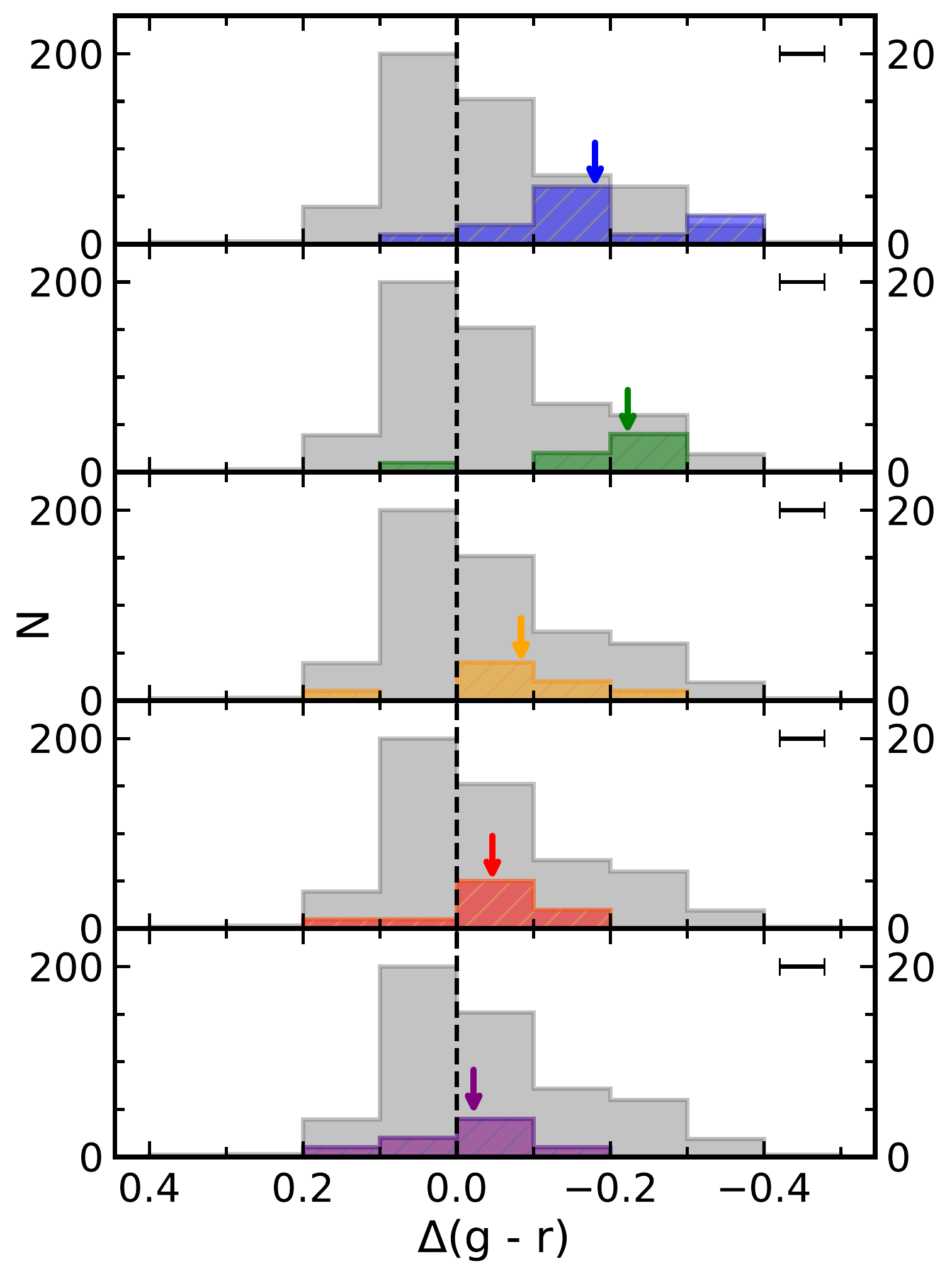}
    \end{subfigure}
    \begin{subfigure}[b]{0.3\textwidth}
        \includegraphics[width=\textwidth]{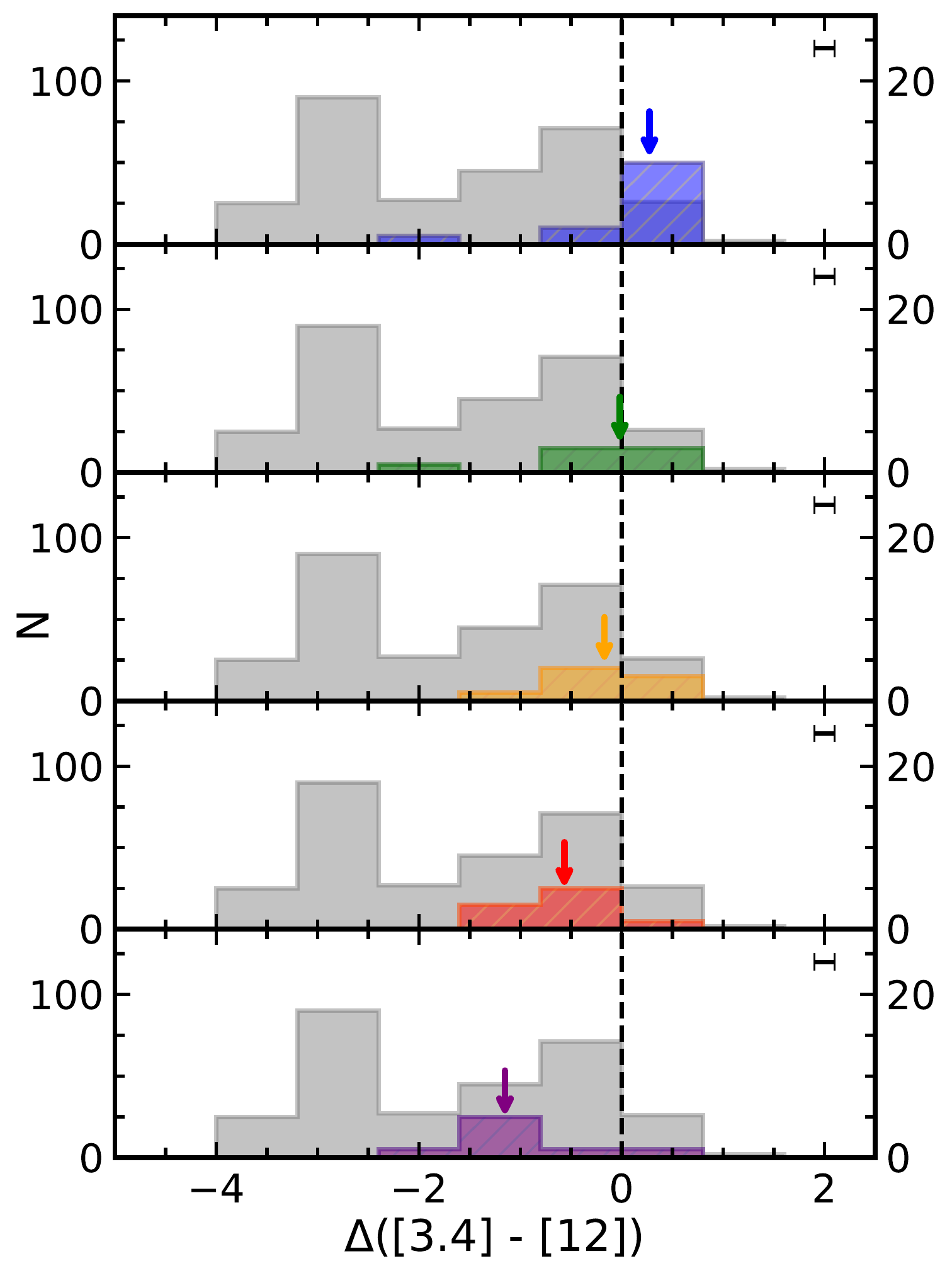}
    \end{subfigure}
\caption{Starburstiness (R\textsubscript{SB}), $\Delta$(\textit{g $-$ r}), $\Delta$([3.4] $-$ [12]) histograms (from left to right). Vertical lines denote the locus of the reference sequence (i.e., star-forming main sequence for R\textsubscript{SB}, red sequence for \textit{g $-$ r}, star-forming sequence for [3.4] $-$ [12]). Certain members, for which cluster membership is given by EVCC, with no defined RPS classifications are plotted together for reference. The arrows represent the median values for each sample. Bin counts were increased by a factor of 5 (10 for the $\Delta$(\textit{g $-$ r}) distribution) for RPS galaxies for a fair comparison with the reference sample. The scale of the original bin counts are labeled on the right hand side of each panel. The representative (either the mean or the median) errors for each parameter are shown in the top right corner of each panel. \label{fig:fig4}}
\end{figure*}

\subsubsection{Star Formation Properties of Galaxies with Different RPS Classes} \label{subsubsec:rpssf}
We examine the star formation properties of galaxies undergoing different stages of ram pressure stripping, using star formation tracers sensitive to a broad range of star-forming timescales. We use SFR\textsubscript{IR}, \textit{g $-$ r}, and \textit{WISE} [3.4] $-$ [12] colors. We plot these star formation tracers against stellar mass in Figure \ref{fig:fig3}. For SFR\textsubscript{IR}, we adopt the main sequence defined in \citet{elbaz07}, which was derived from a best linear fit of star-forming SDSS galaxies at 0.015 $\leq$ \textit{z} $\leq$ 0.1. The solid line represents the main sequence, with the dashed lines representing the lower and upper bounds to the main sequence fit. The optical color \textit{g $-$ r} can also work as a tracer of star formation activity \citep{strateva01, blanton03}. We define a red sequence using the best linear fit of non-RPS class galaxies with M\textsubscript{*} $>$ 1.5$\times$10\textsuperscript{10} M\textsubscript{$\odot$} and \textit{g $-$ r} $>$ 0.58. The mid-infrared [3.4] $-$ [12] color is a tracer of recent star formation activity in a galaxy, up to $\sim$2 Gyr timescales (e.g., \citealt{ko13}). It is also known that 12$\mu$m and 3.4$\mu$m luminosities show strong correlations with SFRs and stellar masses, respectively. As such, [3.4] $-$ [12] color works as a proxy of SFRs per stellar mass (i.e. specific SFRs) \citep{donoso12, li07}. We adopt the peak of the mid-infrared color distribution of star-forming galaxies (i.e., [3.4] $-$ [12] = 1) from \citet{hwang12b} as the star-forming sequence. 

Largely due to small number statistics of the RPS class sample, it is difficult to identify a trend in the star formation activity of RPS class galaxies from Figure \ref{fig:fig3} alone. Overall, galaxies in the Virgo cluster appear to have lower current star formation rates relative to the star-forming main sequence. There are a select few Class 0 and I galaxies found above the main sequence. However, there are also some others that appear to be less star-forming compared to other RPS class galaxies that are far more deficient in HI. We note that the main sequence adopted here was based on a best linear fit of galaxies located at 0.015 $\leq$ \textit{z} $\leq$ 0.1 \citep[see][Figure 18]{elbaz07}, and thus may not be an accurate representation of the main sequence for galaxies at \textit{z} $\lesssim$ 0.01, which is the case for galaxies in the EVCC. As such, we do not use the main sequence as an absolute definition of the star-forming main sequence for the EVCC sample, but rather as a reference sequence for computing starburstiness. In terms of optical color, RPS class galaxies are overall blue relative to the red sequence, with Class IV galaxies noticeably redder in comparison. As for mid-infrared colors, Classes I and II galaxies are found to be distributed around the star-forming sequence, despite undergoing RPS. This seems to contradict the notion given by their total IR-based SFRs, as the two classes overall exhibit low SFRs with respect to the star-forming main sequence. This disparity can be explained by the fact that mid-infrared colors trace longer star-forming timescales than those of total IR-derived star formation rates (up to $\sim$100 Myr; \citealt{kennicuttevans12}). On the other hand, Class III and IV galaxies overall exhibit lower values of mid-infrared colors (primarily corresponding to lower 12$\mu$m emission), which is not surprising considering their HI morphology and range of HI deficiencies. In contrast to the clustering of galaxies around the star-forming sequence, we also identify a passive sequence located around [3.4] $-$ [12] $\sim$ -2.0. According to Figure 4c from \citet{ko13}, where they compare their sample of quiescent red sequence galaxies with single stellar population (SSP) models that include mid-infrared emission from AGB stars, the mid-infrared colors of the passive galaxies are roughly consistent with stellar populations with a mean age of $\gtrsim$5 Gyr. Those with [3.4] $-$ [12] $\lesssim$ -2.0 can be described by SSP models without AGB dust. Such galaxies with relatively old stellar populations are found to lie on the optical red sequence and exhibit much lower star formation rates. 

We note that RPS class galaxies exhibit different ranges in stellar mass, which shows that there is a need to minimize mass effects. We thus choose to examine the distributions of starburstiness, $\Delta$(\textit{g $-$ r}), and $\Delta$([3.4] $-$ [12]). Starburstiness (R\textsubscript{SB} = sSFR/sSFR\textsubscript{MS}; \citealt{elbaz11}) is a measure of the relative excess of the specific star formation rate (sSFR = SFR/M\textsubscript{*}) of a galaxy compared to that of a main sequence star-forming galaxy of the same mass. $\Delta$(\textit{g $-$ r}) is a measure of how blue (i.e., star-forming) a galaxy is relative to a red sequence galaxy of the same mass, with negative values indicating bluer galaxies and positive values redder galaxies. $\Delta$([3.4] $-$ [12]) measures the deficiency in IR color (which is a proxy for deficiency in star formation) of a galaxy relative to a galaxy of the same mass on the star-forming sequence, with positive values signifying galaxies rich in IR color. Using these parameters thus allows a fair comparison of the star formation tracer distributions between different RPS classes.

Figure \ref{fig:fig4} shows the resulting histograms of the star formation tracer distributions - starburstiness, $\Delta$(\textit{g $-$ r}), and $\Delta$([3.4] $-$ [12]) - from left to right. We denote the median of each distribution with downward arrows. We overlay the distribution of each RPS class on top of that of other cluster members with no RPS classification, as reference. In general, the resulting distributions of each star formation activity indicator seem to be consistent with one another. Classes 0 and I seem to make up the actively star-forming end (i.e., higher starburstiness, bluer colors, and higher mid-IR colors) of the cluster population, whereas Classes II and III tend to lie in the range of moderately to passively star-forming. On the other hand, Class IV galaxies appear to be the most quenched in star formation activity for all star formation tracers. We also note the presence of galaxies with lower starburstiness and $\Delta$([3.4] $-$ [12]) than RPS class galaxies, which can be attributed to the sequence of quiescent galaxies observed in the bottom panel of Figure \ref{fig:fig3}. However, the differences between Classes I through III do not show to be very clear, hinting at the limitations of small number statistics. We perform Kolmogorov-Smirnov (K-S) tests to examine the star formation activity distributions in more detail.

We are not able to observe any noticeable enhancement in star formation activity for galaxies undergoing early to active stripping, i.e., Classes I and II. Despite a shift in the median of the $\Delta$(\textit{g $-$ r}) distribution towards bluer colors for Class I relative to that of Class 0, we are not able to reject the hypothesis that the two distributions are drawn from the same population (i.e., the null hypothesis), even at a $>$2$\sigma$ level, for the two classes for any of the star formation activity indicators. Taking into account that the HI deficiency ranges of Class 0 and I galaxies are overall comparable to one another (see Table \ref{tab:tab1}), it is not surprising that the distinction between the two classes is statistically insignificant. Classes I and II also show no statistically significant differences from one another. On the other hand, we are able to reject the null hypothesis for Class 0 and III, and Class 0 and IV at a $>$2$\sigma$ level for all star formation tracers. The results of the K-S test on the $\Delta$([3.4] $-$ [12]) distribution gives $p_{KS}$ $\approx$ 0.001 for Classes 0 and III, and for Classes 0 and IV, indicating a significance of $>$3$\sigma$. Although there seems to be a trend that the distributions of star formation tracers systematically change with stripping classes, such differences are not strongly supported by statistical tests. We suspect that this is largely due to small number statistics, and the fact that we are using integrated photometry to trace the star formation activity in these galaxies. Moreover, we also consider the possibility that if there is any enhancement in star formation induced by ram pressure stripping, it is unlikely to last for a long enough period of time to be traceable by the star formation tracers used in this study. We discuss the implications of our results in further detail in Section \ref{sec:disc}. We observe the most statistically significant differences for RPS classes with large disparities in HI deficiencies, relative HI extent, and HI morphology, such as Class 0 and III, and Class 0 and IV. 

\subsection{Star Formation Properties of Galaxies Classified by HI Mass Fractions and Positions in Phase Space} \label{subsec:higroup}
While we are able to capture the overall quenching of star formation activity with increasing degree of gas stripping with RPS classes, the differences among the stripping classes are not statistically significant. As high-resolution HI imaging is only available for a limited sample of the EVCC, small number statistics do come into play. Moreover, because \citet{yoon17}'s classification is primarily based on HI morphology, we do require a more quantitative classification scheme on top of a larger sample. We thus aim to expand the sample by re-defining new stripping groups using a combination of HI mass fraction (HI-to-stellar mass ratios; M\textsubscript{HI}/M\textsubscript{*}) and location in projected phase space. While \citeauthor{yoon17} categorized galaxies into different stripping classes based on HI properties, and then traced the orbital histories of such galaxies using location in projected phase space, we utilize both HI mass fractions and expected position in projected phase space according to radial infall to define a new classification scheme. We choose to use HI-to-stellar mass ratios as they are not dependent on a separate sample of isolated galaxies, like HI deficiencies are. Moreover, given that HI-to-stellar mass ratios are more commonly used to measure HI gas content, we believe that HI-to-stellar mass ratios would be a better choice if one were to compare the results of this study to those of other studies. We visually inspect all EVCC galaxies with HI detections to identify and remove a total of 11 spheroidal galaxies that may have been subject to morphological quenching \citep{martig09} from the sample. This leads to a total of 519 spiral and irregular galaxies with HI detections in the EVCC, which includes all 48 of \citeauthor{yoon17}'s RPS class galaxies.

\begin{figure*}[ht!]
\centering
    \begin{subfigure}[b]{0.48\textwidth}
        \includegraphics[width=\textwidth]{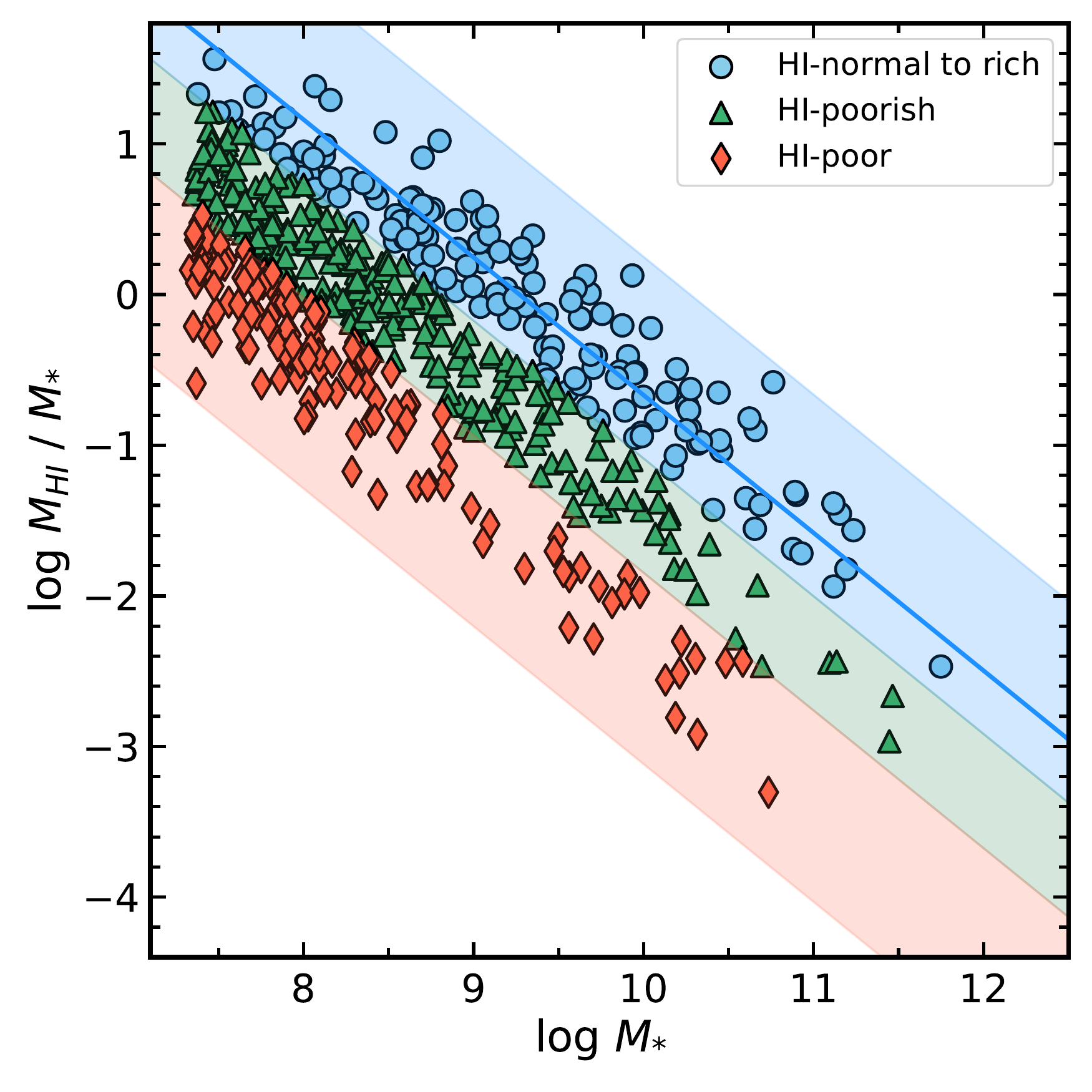}
    \end{subfigure}
    \begin{subfigure}[b]{0.48\textwidth}
        \includegraphics[width=\textwidth]{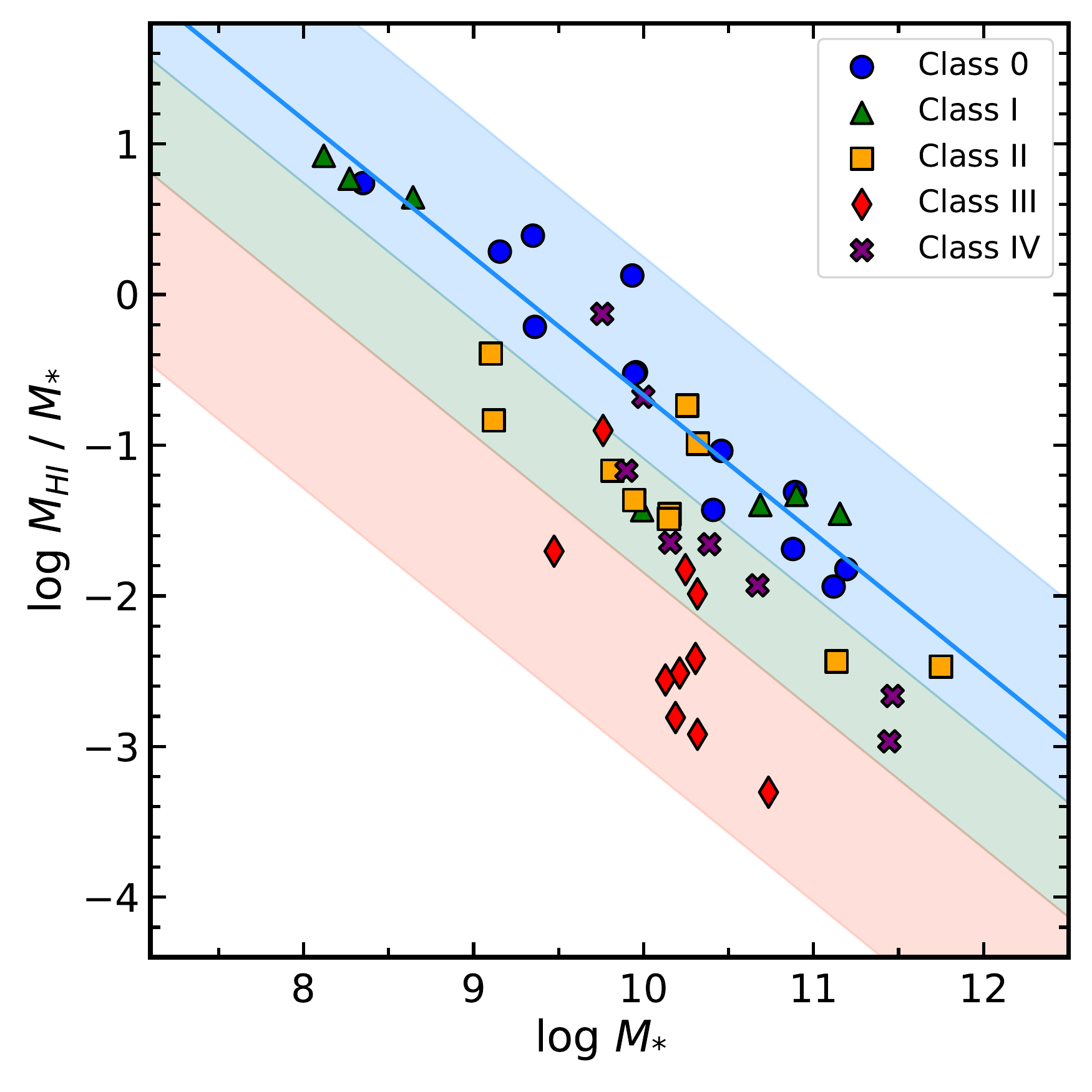}
    \end{subfigure}
\caption{(Left) HI mass fraction as a function of stellar mass, for which each of the shaded regions mark the range of HI mass fractions of each group, as defined by Eqs. \ref{eq:2} through \ref{eq:4}. The blue line is the reference sequence as defined by a best linear fit of Class 0 and I galaxies, defined by Eq. \ref{eq:1}. (Right) HI mass fraction as a function of stellar mass for the RPS class sample, to verify whether the selected ranges of $\Delta$($\log$ M\textsubscript{HI}/M\textsubscript{*}) for each group are consistent with those of the RPS class sample. \label{fig:fig5}}
\end{figure*}

\begin{figure}[htb!]
\centering
\includegraphics[width=0.45\textwidth]{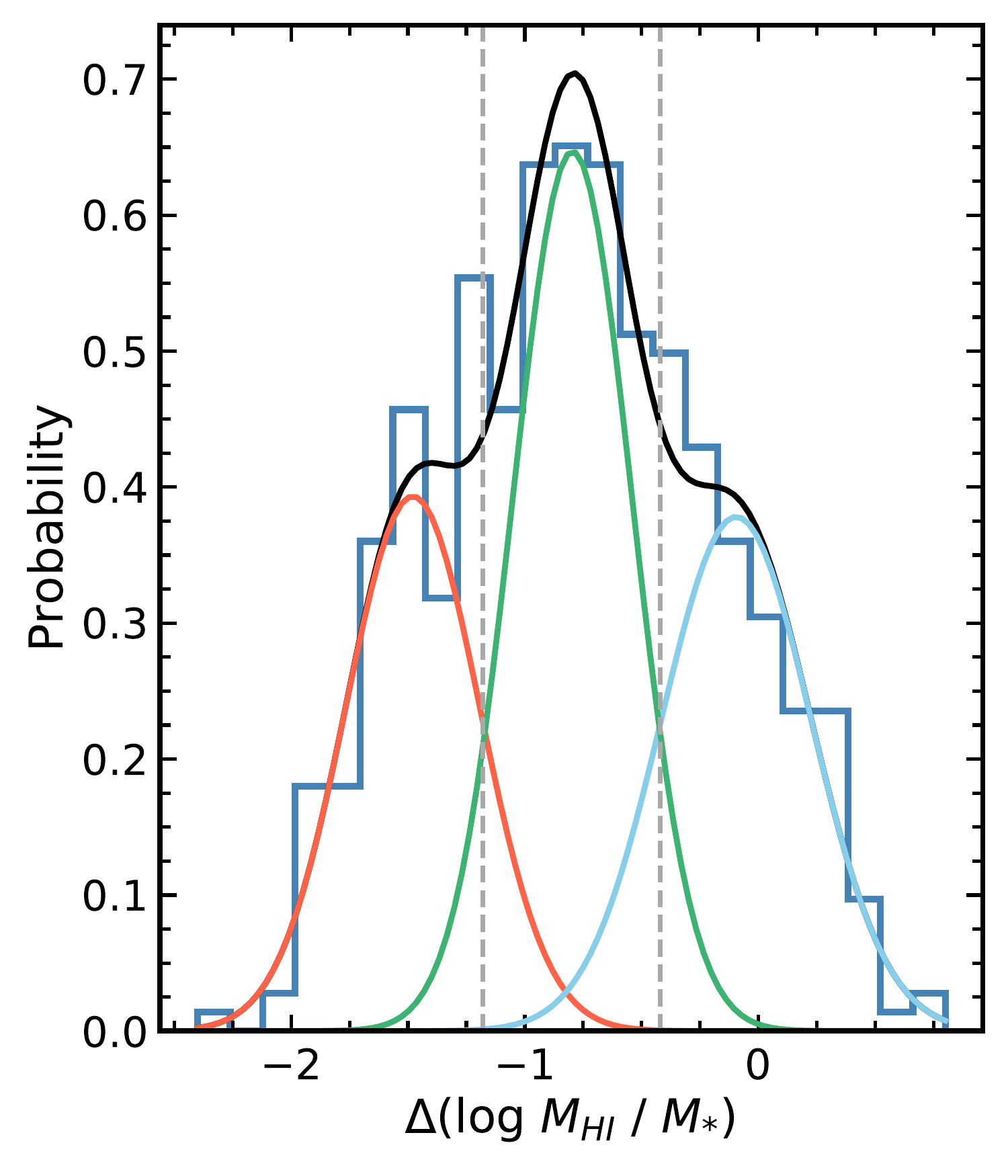}
\caption{$\Delta$($\log$ M\textsubscript{HI}/M\textsubscript{*}) histogram with the resulting Gaussian Mixture Model (GMM) overlaid on top. Each Gaussian component specifies the mean and range of a potential HI group. The blue component represents the HI-normal to rich galaxies, the green represents the HI-poorish galaxies, and the red the HI-poor. The grey dashed lines mark the intersections between the different Gaussian components. \label{fig:fig6}} 
\end{figure}

\subsubsection{Definition of HI Stripping Groups} \label{subsubsec:higroupdef}
To define new HI stripping groups, we first plot the HI mass fraction as a function of stellar mass, and perform a best linear fit of Class 0 and I galaxies to define a reference sequence (Eq. \ref{eq:1}), as shown in Figure \ref{fig:fig5}. Despite differences in HI morphology, Class 0 and I are similar in HI deficiencies (see Table \ref{tab:tab1}) and thus the two classes can be thought to represent the HI-rich end of galaxies in the Virgo cluster. As similarly done in Figure \ref{fig:fig4}, we measure $\Delta$($\log$ M\textsubscript{HI}/M\textsubscript{*}), the deficiency of the HI mass fraction of a galaxy with respect to that of a galaxy on the reference sequence with the same mass. We then use Gaussian Mixture Modeling (GMM) to identify three groups defined by richness in their HI gas content: HI-normal to rich, HI-poorish, and HI-poor. The GMM results are shown in Figure \ref{fig:fig6}. We adopt the intersections of the Gaussian components, i.e., the dashed grey lines in Figure \ref{fig:fig6}, as the boundaries between the different groups. The equations defining the ranges of $\Delta$($\log$ M\textsubscript{HI}/M\textsubscript{*}) for each group are given by Eqs. \ref{eq:2}, \ref{eq:3}, and \ref{eq:4}, denoting HI-normal to rich, HI-poorish, and HI-poor, respectively. 

\begin{figure*}[ht!]
\centering
\includegraphics[width=0.8\textwidth]{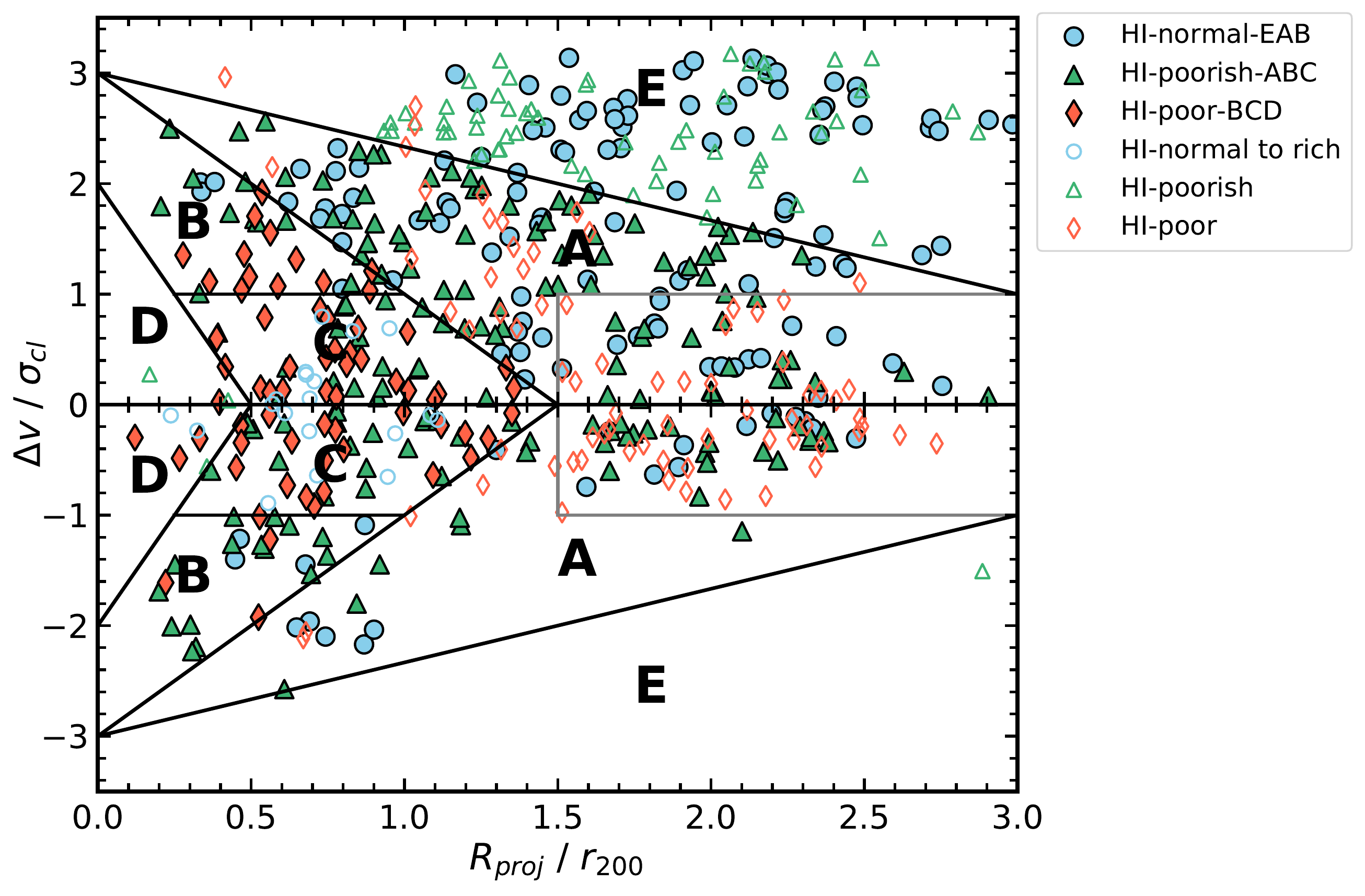}
\caption{Projected phase-space diagram of all newly defined HI stripping groups. The black solid lines indicate the boundaries between different regions of time since infall as shown in Figure \ref{fig:fig2}. The final group candidates are shown as filled color symbols, whereas those that did not satisfy the condition of expected location in phase space are shown as open symbols. Red open symbols (HI-poor galaxies located outside of Regions B through D) residing in the region defined by grey solid lines are selected as potential backsplashing candidates, which are discussed briefly in Section \ref{subsubsec:backsp}. \label{fig:fig7}}
\end{figure*}

\begin{table*}[htb!]
\caption{Categorization of the EVCC into different HI stripping groups \label{tab:tab2}}
\captionsetup{justification=justified}
    \centering
    \begin{tabular}{ccc}
    \toprule
        HI Stripping Group & $\Delta$($\log$ M\textsubscript{HI}/M\textsubscript{*}) & Location in phase space \\
         & (Range) &  \\
        (1) & (2) & (3) \\
    \midrule
        HI-normal-EAB &  Eqn. \ref{eq:2} & Regions E, A, or B \\
        (N = 132) & & \\
        HI-poorish-ABC & Eqn. \ref{eq:3} & Regions A, B, or C \\
        (N = 168) & & \\
        HI-poor-BCD & Eqn. \ref{eq:4} & Regions B, C, or D \\
        (N = 65) & & \\
    \bottomrule
    \end{tabular}
\caption*{(1) HI stripping group, (2) HI mass fraction range, and (3) location in phase space.}
\end{table*}

\begin{align} \label{eq:1}
    \log \frac{M_{HI}}{M_{*}} = -0.915 \log M_{*} + 8.478
\end{align}

\begin{align} \label{eq:2}
\begin{split}
    & \text{HI-normal to rich:} \\
    & \log \frac{M_{HI}}{M_{*}} + 0.915 \log M_{*} \geq (8.478 - 0.42)
\end{split}
\end{align}

\begin{align} \label{eq:3}
\begin{split}
    & \text{HI-poorish:} \\
    & \begin{cases}
        \log \frac{M_{HI}}{M_{*}} + 0.915 \log M_{*} < (8.478 - 0.42) \\
        \log \frac{M_{HI}}{M_{*}} + 0.915 \log M_{*} \geq (8.478 - 1.179)
    \end{cases}
\end{split}
\end{align}

\begin{align} \label{eq:4}
\begin{split}
    & \text{HI-poor:} \\
    & \log \frac{M_{HI}}{M_{*}} + 0.915 \log M_{*} < (8.478 - 1.179)
\end{split}
\end{align}

With these definitions, we classify galaxies in the EVCC with HI detections into a total of three groups, which are visualized in the left panel of Figure \ref{fig:fig5}. On the right, we verify whether the selected ranges of HI mass fractions are consistent with those of the RPS class sample. Despite the presence of some scatter, most of Class 0 and I galaxies are well represented as HI-normal to rich galaxies. The selected range of HI mass fractions for the HI-poorish group is consistent with those of Class II and III galaxies, and we confirm that more than half of the Class III objects are categorized as HI-poor galaxies. With the exception of one galaxy, Class IV galaxies are generally well described as HI-poorish, which is expected given that their HI deficiency range overlaps with that of Class II galaxies (see Table \ref{tab:tab1}). We thus affirm that the selected ranges of $\Delta$($\log$ M\textsubscript{HI}/M\textsubscript{*}) for each group, as given by the results of GMM, are well founded.

To distinguish galaxies at different stages in their orbital histories, we employ the use of projected phase space as a second means of categorizing galaxies into HI stripping groups, as shown in Figure \ref{fig:fig7}. We utilize the same definitions of regions in phase space as shown in the left panel of Figure \ref{fig:fig2}. Galaxies that are likely to be on their first infall are expected to be found in Region A. For those that are approaching their first core crossing, or have already passed and are falling out of the core, they are likely to be found in Region B. Galaxies that have gone through multiple core crossings will eventually settle and virialize within the cluster potential, finding themselves at low velocities and clustercentric distances, such as Regions C or D. However, as confirmed by \citet{yoon17}, it is possible that some galaxies may reach further out after outfall, otherwise known as backsplashing galaxies. We briefly examine the possibility of finding backsplashing candidates with our group categorization in Section \ref{subsubsec:backsp}. Results from \citet{jaffe15} suggest that star-forming galaxies relatively rich in HI gas content are found to be recent infallers, whereas passive galaxies with low gas content are found in the virialized zone. This suggests that galaxies are expected to lose their gas and become increasingly quenched in star formation activity as they fall into the cluster core. As for galaxies that are located at relatively high velocities and large clustercentric distances, they are less subject to the cluster potential and thus are expected to retain a comparable amount of HI mass as that of a field galaxy. We may find such galaxies in Region E. However, we stress that the HI-normal to rich galaxies located in Region E may not be an ideal control sample of galaxies outside of the cluster potential well, as those presumed to be background galaxies were excluded from the EVCC with a radial velocity cut, as mentioned in Section \ref{sec:data}. Keeping the typical infall trajectory along with the expected trend in star formation activity and gas content in mind, we finalize our HI group sample based on their expected position in phase space, in combination with their HI mass fractions, resulting in a total of 365 galaxies. The final HI group sample is presented in Table \ref{tab:tab2}, labeled as HI-normal-EAB, HI-poorish-ABC, and HI-poor-BCD. While there were many other possible combinations of the two parameters we could have used, our classification scheme was chosen such that galaxies of each group fell in the expected region of phase space as those of corresponding RPS classes (see Figure \ref{fig:fig2}). For example, HI-normal-EAB galaxies were chosen such that they roughly lie in the same region as Class 0 and I galaxies. HI-poorish-AB were meant to represent Class II galaxies, whereas HI-poor-BCD represent Class III. Amongst the final sample, there is a total of 43 galaxies (out of 48) that comprise the RPS class sample in \citet{yoon17}. There are 19 galaxies categorized as HI-normal-EAB, which consist of RPS Classes 0 - II, and IV. The HI-poorish-ABC group includes 17 RPS class galaxies, which consist of a mixture of Classes I - IV. The HI-poor-BCD group consists of 6 RPS Class III galaxies. The remaining galaxy is one of the confirmed backsplashing candidates in \citet{yoon17}, which is included in our analysis of potential backsplashing candidates in Section \ref{subsubsec:backsp}. 

\begin{figure*}[htb!]
\centering
\includegraphics[width=0.8\textwidth]{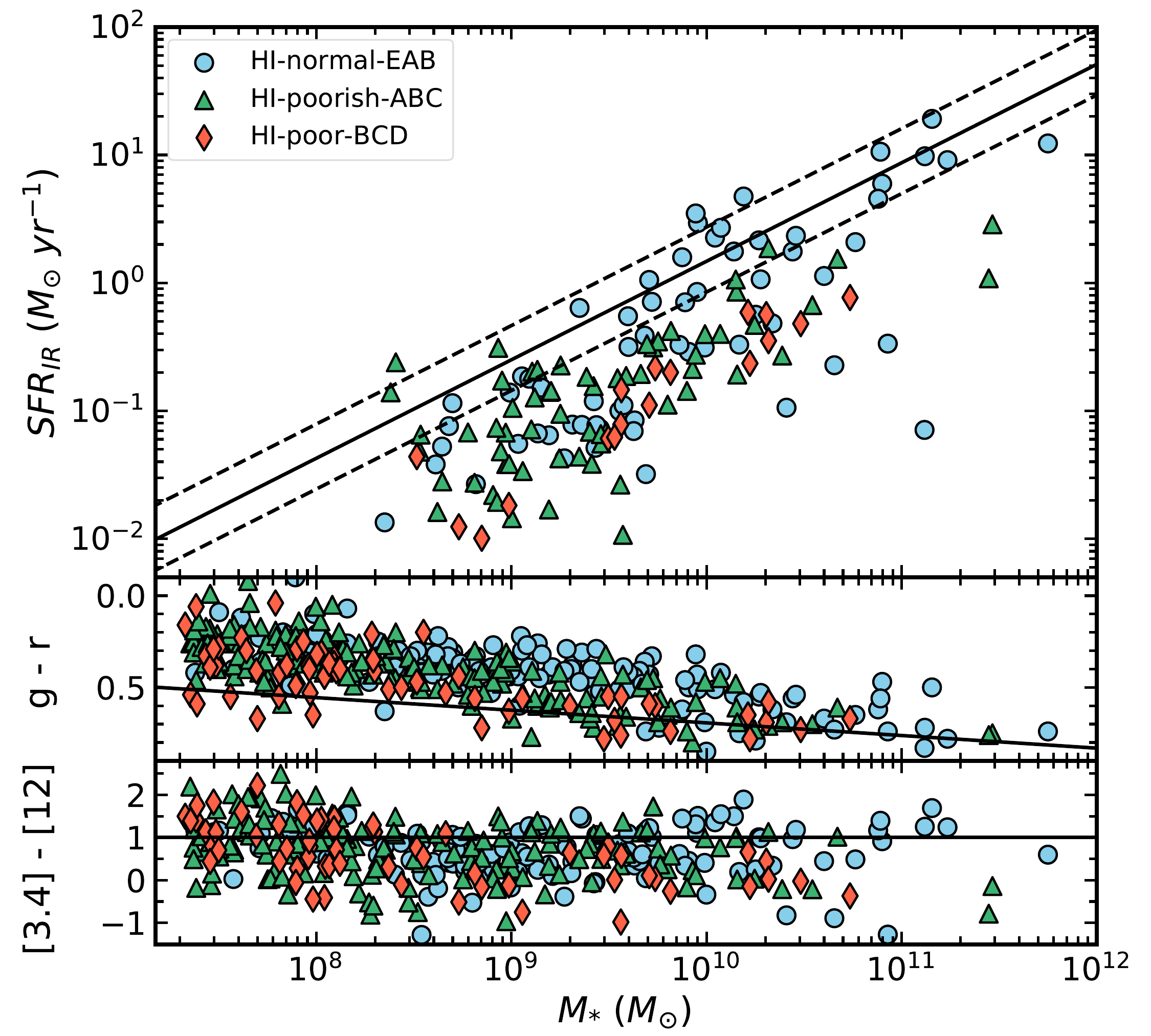}
\caption{Star formation rates derived from total IR luminosities, optical, and mid-infrared colors (from top to bottom) as a function of stellar mass for the EVCC sample with newly defined HI stripping groups. The reference sequences shown here are identical to the ones used in Figure \ref{fig:fig3}. \label{fig:fig8}}
\end{figure*}

\begin{figure*}[htb!]
\centering
    \begin{subfigure}[b]{0.3\textwidth}
        \includegraphics[width=\textwidth]{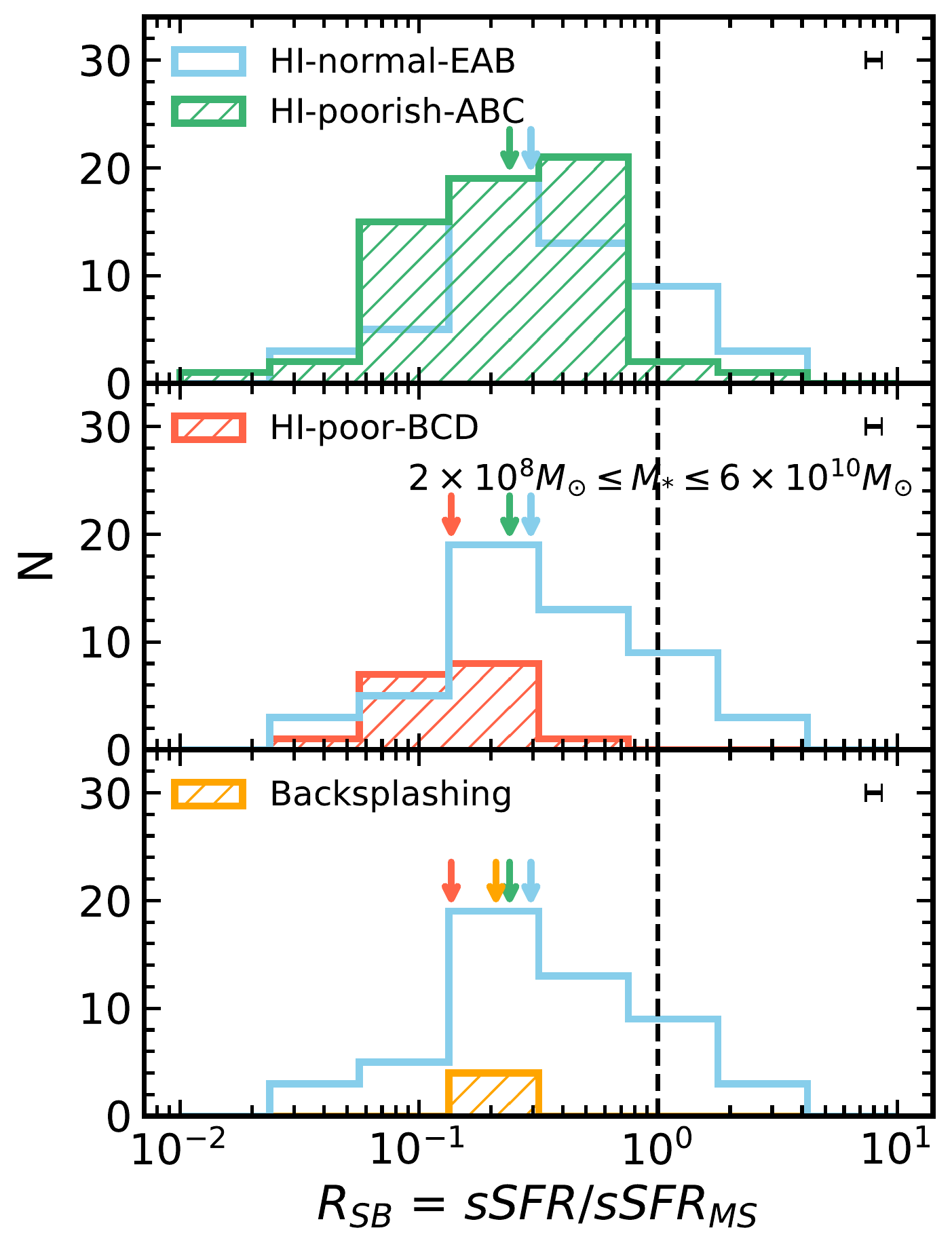}
    \end{subfigure}
    \begin{subfigure}[b]{0.3\textwidth}
        \includegraphics[width=\textwidth]{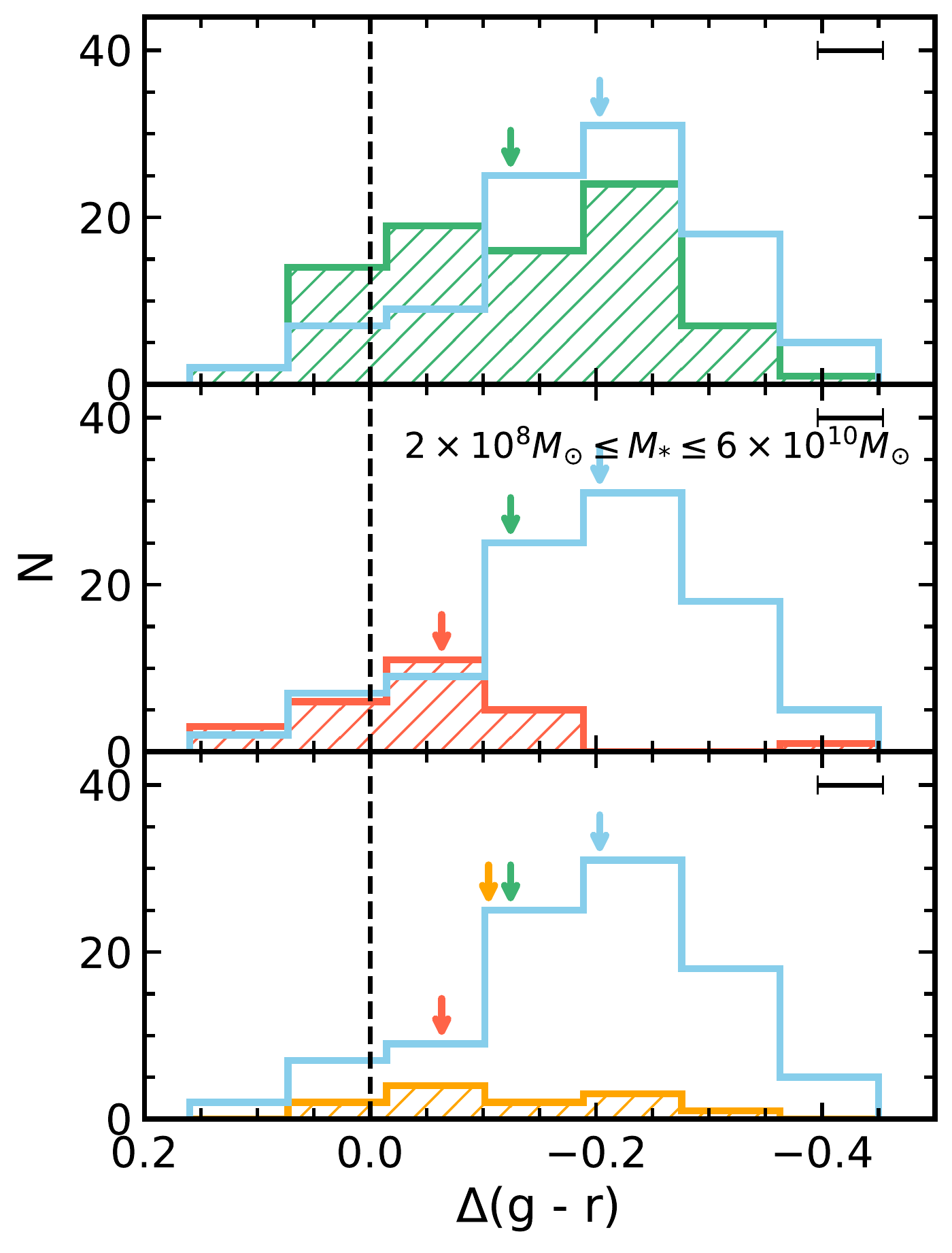}
    \end{subfigure}
    \begin{subfigure}[b]{0.3\textwidth}
        \includegraphics[width=\textwidth]{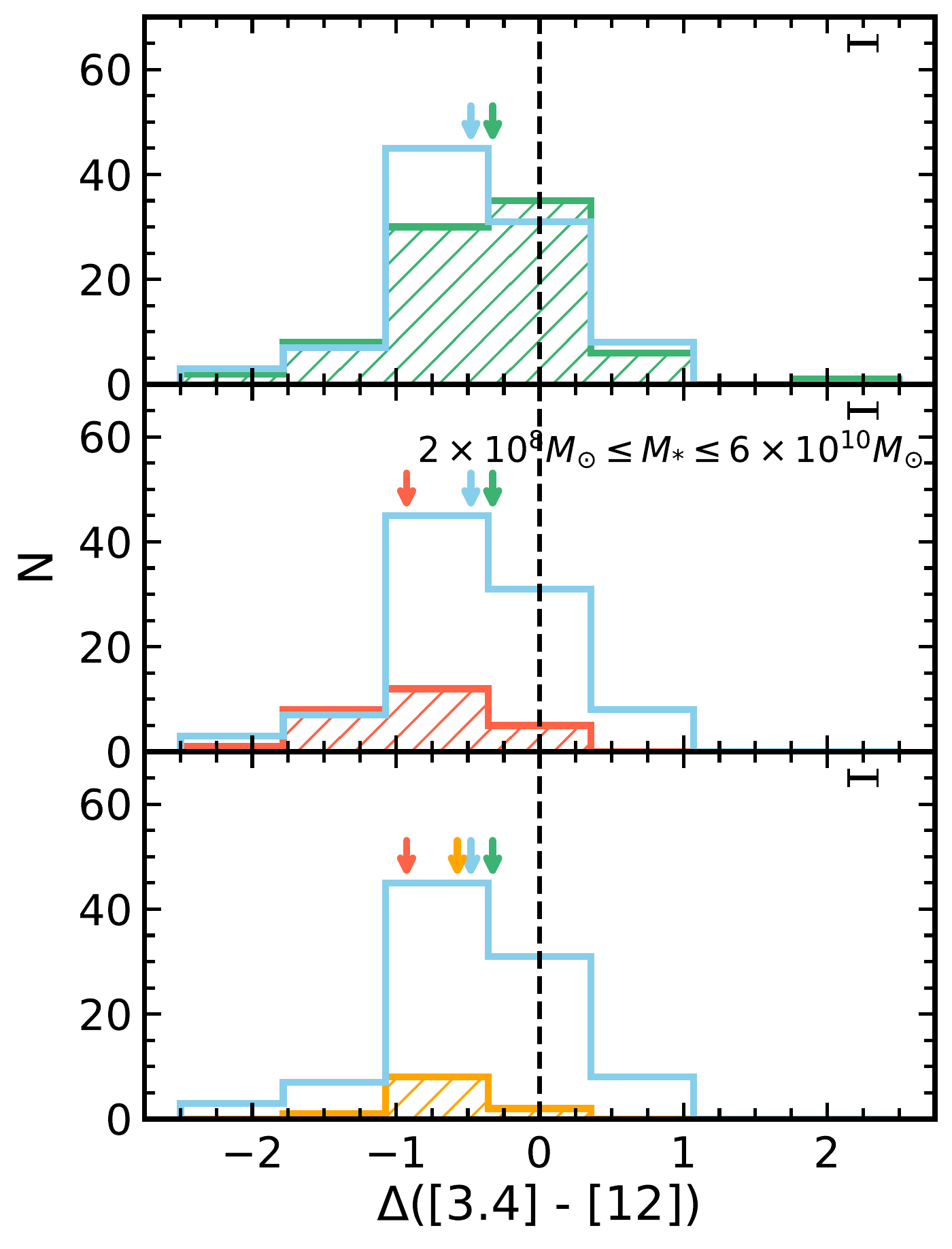}
    \end{subfigure}
\caption{Starburstiness (R\textsubscript{SB}), $\Delta$(\textit{g $-$ r}), $\Delta$([3.4] $-$ [12]) histograms (from left to right) for new HI stripping groups. The downward arrows mark the median of each distribution. The vertical lines here are the same lines defined in Figure \ref{fig:fig4}. Stellar mass range of the subsamples used are displayed in the middle panel of each histogram. Additionally, we show the distribution of potential backsplashing candidates in the bottom panel. \label{fig:fig9}}
\end{figure*}

When plotting the projected phase-space diagram, we take on the same assumptions as done in \citet{yoon17}. Here, we assume that the center of the Virgo cluster is at M87 \citep{bohringer94}. We normalize the projected clustercentric distance by $r_{200}$ and the line-of-sight clustercentric velocity by the cluster dispersion, $\sigma_{cl}$. The clustercentric velocity is calculated using the difference between a galaxy's velocity along the line-of-sight and the mean radial velocity of the M87 subgroup, 1088 km s\textsuperscript{-1} (Cluster A; \citealt{mei07}). We adopt values of $r_{200}$ = 1.55 Mpc \citep{mclaugh99, ferrarese12} and $\sigma_{cl}$ = 593 km s\textsuperscript{-1} \citep{mei07}.

\subsubsection{Star Formation Properties of New HI Stripping Groups} \label{subsubsec:higroupsf}
We examine the star formation properties of the newly defined HI stripping groups, using a larger sample. Using the same star formation tracers as before, we show our results in Figure \ref{fig:fig8}. As seen similarly with Figure \ref{fig:fig3}, we are not able to observe a specific trend in the HI stripping groups, especially given that a disparity in the range of stellar mass is apparent between them. For example, a large fraction of HI-poorish-ABC and HI-poor-BCD galaxies appear to dominate the low-mass end, whereas the high-mass end seems to be increasingly dominated by HI-normal-EAB galaxies. In fact, there are some HI-normal-EAB galaxies found below the star-forming main sequence in the low-mass end. Upon investigating them, we find that they belong to galaxy groups, which suggests that they are undergoing pre-processing prior to cluster infall \citep{kourkchi17}. To minimize mass effects, we perform the same in-depth analysis as done in Section \ref{subsubsec:rpssf}, where we minimize biases due to mass effects by plotting the histograms of starburstiness, $\Delta$(\textit{g $-$ r}), and $\Delta$([3.4] $-$ [12]). Moreover, for a fair comparison between the distributions of different star formation tracers, we perform a stellar mass cut of 2$\times$10\textsuperscript{8} M\textsubscript{$\odot$} $\leq$ M\textsubscript{*} $\leq$ 6$\times$10\textsuperscript{10} M\textsubscript{$\odot$}, which is roughly the stellar mass range of galaxies with measured SFR\textsubscript{IR} (as plotted in the top panel of Figure \ref{fig:fig8}), excluding a select few outliers on the high-mass end.

The resulting histograms are shown in Figure \ref{fig:fig9}. The starburstiness distribution shows that the median values are similar for HI-normal-EAB and HI-poorish-ABC groups, while the median of the HI-poor-BCD group is shifted towards lower values of R\textsubscript{SB}. The K-S tests reveal that all 3 groups show differences from one another on the level of $>$2$\sigma$. We believe the lack of a highly significant discrepancy between the 3 groups may be due to the fact that the \textit{WISE} 22$\mu$m detection limit is not deep enough to result in a large sample of galaxies compared to those of other star formation tracers. On the other hand, we see a much clearer trend in the change in star formation activity for $\Delta$(\textit{g $-$ r}); the median $\Delta$(\textit{g $-$ r}) increases towards redder colors with decreasing HI gas content. Our K-S test results suggest that we can reject the null hypothesis at a $>$3$\sigma$ level for the $\Delta$(\textit{g $-$ r}) distribution for all 3 groups. In contrast, we observe slightly different results with the $\Delta$([3.4] $-$ [12]) distribution. There is a shift in the median $\Delta$([3.4] $-$ [12]) towards larger values, which can suggest an enhancement of star formation activity in HI-poorish-ABC galaxies, relative to that of HI-normal-EAB galaxies. We then observe a quenching of the star formation activity of HI-poor-BCD galaxies with the median shifted towards negative values of $\Delta$([3.4] $-$ [12]). However, the supposed enhancement of star formation activity observed by the shift in the median is not expected to be meaningful as the K-S test does not yield significant results and the difference in the median values are only slightly out of the range of error. On the other hand, we observe a statistically significant difference between HI-normal-EAB and HI-poor-BCD galaxies, and HI-poorish-ABC and HI-poor-BCD galaxies with a $p$-value $<$ 0.004 for both. In general, our results suggest that there is no statistically significant difference between HI-normal-EAB and HI-poorish-ABC galaxies. However, we are able to reject the null hypothesis at a $>$3$\sigma$ level for the optical color distribution. We also observe a significant difference between HI-poorish-ABC and HI-poor-BCD galaxies for all star formation tracers. Overall, we are able to capture the quenching of star formation activity with increasing deficiency in HI gas content with a larger sample compared to our analysis in Section \ref{subsubsec:rpssf}.

\section{Discussion} \label{sec:disc}
\subsection{Locally Enhanced Star Formation in Galaxies Undergoing Gas Stripping} \label{subsec:localsf}
We examine the star formation activity of galaxies undergoing ram pressure stripping in the Virgo cluster using various tracers. We initially perform a follow-up study of \citet{yoon17}, given the availability of high-resolution HI-imaging thanks to the VIVA survey \citep{chung09}. While our results confirm the previously known trend of star formation quenching with increasing degree of ram pressure stripping \citep{jaffe15, yoon17}, the difference between the distributions of galaxies undergoing early through post stripping is not statistically significant. However, we do observe a statistically significant distinction between Class 0 and III, and Class 0 and IV galaxies. 

\subsubsection{Implications from Observational Studies Based on the Virgo Cluster}\label{subsubsec:obssf}
The lack of a conspicuous enhancement in star formation activity is not too surprising given the results of previous studies on the Virgo cluster.  \citet{kk04b} utilize the spatial distributions of H$\alpha$ and $R$-band emission to investigate the environmental effects on the star formation activity on 52 galaxies in the Virgo cluster, among which 39 of them are RPS class galaxies. Most of the H$\alpha$-truncated spirals show signs of having been subject to ram pressure stripping, where many of them exhibit normal to slightly enhanced star formation rates in the inner galaxy disks. While they find asymmetric enhancements in star formation at the outer edge of the H$\alpha$ disk for a few galaxies, including NGC 4654 (Class II) and NGC 4405 (Class III), which is indicative of ongoing ICM pressure, these enhancements are at most globally modest and local. They instead find that galaxies with the largest global enhancement in star formation are those that have undergone low-velocity tidal interactions and/or gas accretion. An example of such a galaxy in their sample is NGC 4299 (Class I), which shows evidence of undergoing both RPS and tidal interactions due to its pair galaxy NGC 4294 (Class I). Furthermore, pixel-by-pixel analyses of FUV images by \citet{boissier12} show little to no star formation activity in the stripped HI gas of a select few Class I and II galaxies, with NGC 4522 (Class II) being an exception. However, the authors come to the conclusion that despite the presence of a mild enhancement in star formation, it is unlikely to affect the global star formation activity and thus will not be discernible when considering integrated quantities or large samples of galaxies as a whole. As RPS class galaxies show to be undergoing quenching on a global scale as a result of ongoing gas stripping, any local enhancement in star formation is unlikely to be discernible with integrated photometry. This explains why, on top of small number statistics, the differences in star formation activity between different RPS classes are not statistically significant. 

On the other hand, we observe a significant difference on a $>$3$\sigma$ level between Classes 0 and III, and Classes 0 and IV. Use of integrated photometry works particularly well in cases where galaxies showing large disparities in HI gas content and star formation activity are compared with one another. Moreover, \citet{yoon17} suggest that Class IV galaxies are likely to have undergone starvation \citep{larson80}, given that they show gas depletion at all radii in the disk instead of the expected asymmetry arising from ram pressure stripping in their HI morphology. Such galaxies would have their star formation quenched throughout the disk, which would be apparent when traced with integrated properties. 

While enhancements in star formation may be prominent for galaxies undergoing gas stripping in other galaxy clusters (e.g., \citealt{vulcani18, ramatsoku19}), it is evident that the data used in this study do not show such a global enhancement for galaxies in the Virgo cluster. If one intends to capture the local enhancement of star formation in Virgo galaxies, spatially resolved data would be necessary, as done in \citet{kk04b}. Results of \citet{leerps17} further prove the advantage of using spatially resolved data - they confirm strong enhancement in both far-ultraviolet (FUV) and H$\alpha$ along CO compression within the stellar disk for three Class II galaxies, which support our hypothesis that galaxies undergoing active stripping are likely to experience some sort of star formation enhancement. 

In addition, \citet{kenney14} suggests that whereas galaxies with one-sided tails of young stellar knots are often observed in more distant and massive (i.e., denser ICM and higher peak ram and thermal pressures) clusters such as the Coma cluster, the Virgo cluster itself is not massive enough to completely strip spiral galaxies. However, such features can be observed in lower mass galaxies such as dwarf irregulars, as \citet{kenney14} show with their analysis of IC 3418. IC 3418 was not included in the EVCC, and moreover has been shown to have weak HI \citep{chung09, kenney14} and mid-infrared emission (checked with \textit{WISE} 12$\mu$m and 22$\mu$m images). Thus, our overall results are unlikely to have changed significantly even if we had included dwarf irregular galaxies such as IC 3418, given our choice of star formation tracers. 

\subsubsection{Impact of RPS on the Global Star Formation Activity of Galaxies as Predicted by Simulations}\label{subsubsec:simsf}

Results of previous observational studies, along with ours, suggest that a global enhancement in star formation triggered by RPS is rather mild and challenging to detect when considering integrated quantities or large samples of galaxies as a whole. Studies such as \citet{kenney14} suggest that the degree of enhancement depends on a variety of factors, such as the stellar mass of the galaxy, total mass of the host galaxy cluster, density of the ICM, etc. Moreover, recent simulation studies (e.g., \citealt{oman16, rhee17}) tracing the trajectories of infalling cluster galaxies in phase space have shown that the relative velocity of galaxies, along with the ambient ICM density, changes with the galaxy's location with respect to the cluster center. As such, the impact of RPS on the star formation activity is not a trivial matter, given that it is also closely coupled with the orbital history of the galaxy.

Disagreements with regards to the degree or even the possibility of global enhancement are also found in simulation studies, given the strong dependency on initial conditions. Recent RPS simulations such as \citet{bekki14} and \citet{steinhauser16} suggest that global enhancement of star formation induced by ram pressure is modest at best (e.g., not on the order of a factor of $\sim$10). Furthermore, despite disagreements with one another in the specific conditions for enhancement to occur, both \citet{bekki14} and \citet{steinhauser16} show that global enhancement only takes place under rigorous conditions. As such, the number of galaxies undergoing enhancement likely make up a small fraction of the cluster population, making it less distinguishable when considering the cluster population as a whole.

Just as galaxies undergoing even a modest degree of global enhancement can be rare in galaxy clusters, as insinuated by recent simulation studies, it is also worthwhile exploring the possibility that the enhancement timescale is rather short, which may bring the choice of star formation tracers used in this study into question. However, there is no clear consensus with regards to the timescale over which enhancement lasts as reported by simulation studies. Despite this, one could nevertheless try to constrain the time period for which the enhancement stage can be best captured. Given that the compression of interstellar gas due to ram pressure is expected to be strongest at orbital pericenter (e.g., \citealt{vollmer09, bekki14}), it would be reasonable to assume that a modest starburst could occur around this stage of a galaxy's orbital history. Using models based on galaxies in the Virgo cluster, \citet{vollmer09} claims that the time window during which one could identify perturbations due to RPS around peak ram pressure is $\sim$300 Myr. With this in mind, RPS itself is expected to occur on a relatively short timescale of 100 Myr (e.g., \citealt{roediger09}). SFRs based on TIR luminosities are sensitive to timescales of $\sim$100 Myr \citep{kennicuttevans12}, which implies that our choice of star formation tracers was appropriate. However, it is entirely possible that any enhancement due to RPS may occur for even shorter timescales, such that it would be more advantageous to use star formation tracers sensitive to even shorter timescales ($\sim$10 Myr), such as H$\alpha$. Barring that possibility, our choice of star formation tracers do not seem to be of an issue in terms of sensitivity to star formation timescales.

It is also worth noting that previous studies have found that while RPS can enhance the global SFR for albeit a short period of time, such changes were not discernible in galaxy colors. For example, \citet{fujita99} study the change in the SFR along with a galaxy's optical \textit{B $-$ V} color to find that while the SFR increases by at most a factor of 2, the color stays overall the same. The authors suggest that ram pressure compression does not induce observable star formation activity as the intrinsic scatters of color and luminosity among galaxies are larger. In fact, after stripping occurs, the color of their model galaxy rapidly turns red. \citet{kapferer09} also report similar results, in that despite significant enhancement of the overall SFR due to ram pressure, a blue star-forming disk galaxy is expected to transform into a red, passively evolving system under $<$250 Myr. \citet{steinhauser16} studied the SDSS \textit{u $-$ i} color evolution of their model galaxies to find similar results as well, with the rate of evolution towards redder colors dependent on the strength of ram pressure. As such, while it is feasible to trace the overall quenching of star formation activity with galaxy colors, results of previous studies seem to suggest that galaxy colors are not sensitive to short-term enhancements of the SFR.

\subsection{Depletion of HI Gas Content According to Time Since Infall in Phase Space} \label{subsec:ppshi}
As the RPS classification given in \citet{yoon17} is limited to galaxies with high-resolution HI imaging, we build a larger sample of galaxies undergoing ram pressure stripping in the cluster by using a combination of HI mass fractions and positions in projected phase space. We examine a total of three different groups of galaxies in the EVCC, with the final sample totaling to 365: 1) HI-normal-EAB: HI-normal to rich galaxies located at large clustercentric distances or high velocities, 2) HI-poorish-ABC: HI-poorish galaxies likely to be on their first infall into the cluster core, and 3) HI-poor-BCD: HI-poor galaxies that are in the process of virializing within the cluster core.

\subsubsection{Overall Quenching of Star Formation with Decreasing HI Gas Content and Increasing Time Since Infall} \label{subsubsec:hiquench}
Given the disparity in HI gas content and location in projected phase space, one may expect to see relatively quenched star formation in HI-poorish-ABC galaxies compared to that of the HI-normal-EAB galaxies. However, as mentioned in Section \ref{subsubsec:higroupsf}, the two groups are not necessarily distinct from one another. One possible explanation for the lack of a significant discrepancy is that while it is expected for HI-poorish-ABC galaxies to become quenched as they infall into the cluster core, the number of galaxies that have actually begun quenching may not comprise of a large fraction of the group. In fact, while HI-poorish-ABC galaxies are likely to have begun losing their gas due to the ICM, the effects of gas loss could have yet to be reflected in the tracers of star formation activity used in this study. If the latter is true, then it is not surprising that the star formation activity of HI-poorish-ABC and HI-normal-EAB galaxies are hardly distinguishable from one another. Simulation results reported in \citet{oman16} suggest that star formation quenching occurs after a delay time of typically 3.5 - 5 Gyr, measured from the first crossing of 2.5 $r_{vir}$ ($\sim$3.4 $r_{200}$), which can be thought of as the first time the galaxy enters the cluster. This onset of quenching usually corresponds to times near or shortly after first pericentric approach. As a significant fraction of HI-poorish-ABC galaxies still have yet to make their first pericentric passage, it is likely that they have yet to have their star formation quenched, which can explain the lack of discrepancy in their star formation activity with that of HI-normal-EAB galaxies. On the other hand, we are able to confirm with a $>$2$\sigma$ confidence that HI-poor-BCD galaxies are likely to be at a different stage in their orbital histories and thus are a distinct population of galaxies within the Virgo cluster, compared to HI-normal-EAB and HI-poorish-ABC galaxies.

While we defined sharp boundaries between different regions in projected phase space, categorized according to time since infall as shown by the simulation results of \citet{rhee17}, we stress that the locus of these boundary lines are not to be taken as absolute dividing lines between distinct populations of galaxies in phase space. They are merely to be used as a means of identifying galaxies likely to be undergoing ram pressure stripping given their HI gas fractions in projected phase space. To confirm that our results are not significantly affected by the locus of the boundary lines in phase space, we perform tests by making different modifications to the boundaries. For example, we have tried lowering the upper limit of the boundary line of Region E from $\Delta v/\sigma_{cl}$ = 3 to 2.5,  which led to a decrease in the selection of HI-poorish-ABC galaxies. We also tightened the selection of HI-poor-BCD galaxies by limiting Regions B and C to lower velocities and clustercentric distances. In spite of these alterations, there were no significant changes observable in the star formation distributions. We thus confirm that our results are not significantly changed even if we make different modifications to the boundaries of different regions in phase space.

\subsubsection{Potential Backsplashing Population} \label{subsubsec:backsp}
We also examine a potential backsplashing population of galaxies in the Virgo cluster. The existence of a backsplashing population in the outskirts of galaxy clusters has long been discussed since \citet{balogh00} and \citet{mamon04}. After the first core crossing, galaxies may travel out as far as 2.5 times the virial radius of the cluster \citep{gill05}. Such galaxies are then expected to make another subsequent infall into the cluster core, eventually joining the galaxies that are virializing in the cluster core. We select a potential group of backsplashers by choosing HI-poor galaxies residing in the region $\abs{\Delta v}/\sigma_{cl} \leq$ 1 and $R_{proj}/r_{200} \geq$ 1.5, resulting in a total of 49 galaxies (refer to Figure \ref{fig:fig7}). Amongst the selected candidates, there is one Class III galaxy present, NGC 4064, which was confirmed as a backsplasher in \citet{yoon17}. \citeauthor{yoon17} identified a total of four backsplashing galaxies in projected phase space, of which NGC 4457 is classified as HI-poorish-ABC; NGC 4580 and NGC 4293 are HI-poor-BCD and do not fall within the selected region of backsplashing candidates. The confirmed backsplashing candidates in \citet{yoon17} are Class III galaxies located near $\Delta v/\sigma_{cl} \sim$ 0 and at roughly 1.2 - 1.7 $r_{200}$ in Figure \ref{fig:fig2}. Compared to these galaxies, we have chosen a sample of backsplashing candidates that are overall travelling at relatively larger velocities with respect to the cluster, and some are found even further out to the outskirts around 2.5 - 3 $r_{200}$. We examine the star formation activity of these candidates in comparison to the three HI stripping groups. We show the resulting distributions of each star formation tracer in the bottom panels of Figure \ref{fig:fig9}. We perform the same stellar mass cut for each star formation tracer as described in Section \ref{subsubsec:higroupsf} for consistency. 

Overall, the selected backsplashing group seems to represent an intermediate population between HI-poorish-ABC and HI-poor-BCD galaxies. Taking into account that backsplashing may occur past the cluster's virial radius after first core passage, the trend seen in Figure \ref{fig:fig9} seems to make sense. Although the trend seen in the starburstiness distribution is unclear due to the limited number of galaxies with SFR estimates (i.e., \textit{WISE} 22$\mu$m detections), we are able to observe increased star formation quenching as a galaxy makes multiple pericentric approaches within the Virgo cluster. If they are true backsplashers, we surmise that they will eventually join the HI-poor-BCD galaxies near the cluster core, and become passive as they continue to lose their gas. We also consider the possibility that these galaxies may not be a true backsplashing population, and may very well have lost their gas through means other than ram pressure stripping. When galaxies are accreted into a nearby galaxy cluster, some may fall in as an isolated galaxy, but it is also likely that a significant fraction of them may fall in as a group. Simulation results of \citet{mcgee09} show that a 10\textsuperscript{14.5} \textit{h}\textsuperscript{-1} M\textsubscript{$\odot$} cluster at z = 0 has accreted about 40$\%$ of its galaxies (M\textsubscript{*} $>$ 10\textsuperscript{9} \textit{h}\textsuperscript{-1} M\textsubscript{$\odot$}) from groups with masses greater than 10\textsuperscript{13} \textit{h}\textsuperscript{-1} M\textsubscript{$\odot$}. Prior to infall, galaxies in groups are likely to have their star formation quenched due to mechanisms such as galaxy-galaxy interactions and galaxy-group tidal interactions, a phenomenon known as ``pre-processing" \citep{fujita04}. These galaxies thus may appear to have lost even more gas with respect to their location in projected phase space. Taking into account the presence of filamentary structures such as the Leo II A and B filaments \citep{kim16} around the vicinity of the Virgo cluster, pre-processing is definitely a possibility that we cannot rule out when it comes to interpreting the gas content of galaxies in projected phase space.

\section{Summary and Conclusions\label{sec:con}}
We investigate the star formation properties of galaxies undergoing ram pressure stripping in the Virgo cluster, building upon the analysis of the orbital histories of VIVA galaxies with high-resolution HI imaging as reported in \citet{yoon17}. We use starburstiness, optical, and mid-infrared colors to examine the star formation activity of galaxies undergoing different stages of ram pressure stripping. Due to the limited number of galaxies with high-resolution HI imaging, we extend our study to the EVCC sample using HI mass fraction and position in projected phase space to categorize galaxies into different stages of stripping. We summarize our main results below. 

\begin{enumerate}
    \item We identify a trend of decreasing star formation activity with increasing degree of stripping, as confirmed by previous studies. However, we do not find strong evidence for star formation enhancement in galaxies undergoing early through active stripping. We attribute the lack of a strong discernment towards the insensitivity of integrated photometry towards local star formation enhancements and the possibility of a relatively short enhancement stage.
    \item To make up for the shortcomings of small number statistics, we extend our study towards the EVCC sample, using HI mass fractions and position in projected phase space. We identify three HI stripping groups - HI-normal to rich, HI-poorish, and HI-poor - represented by different ranges of HI mass fractions. We then perform a second means of categorization by expected location in projected phase space according to radial infall into the cluster core.
    \item While we are not able to observe a strong discernment in the star formation properties between the HI-normal-EAB and HI-poorish-ABC groups, the HI-poor-BCD group shows to be clearly quenched in star formation with respect to the other two groups. We attribute the lack of an obvious distinction between HI-normal-EAB and HI-poorish-ABC galaxies to the late onset of star formation quenching. 
    \item We investigate a potential backsplashing population, selected from HI-poor galaxies located at low velocities and large clustercentric distances beyond the virial radius of the cluster. Distributions of their star formation activity show that they may represent an intermediate population between HI-poorish-ABC and HI-poor-BCD galaxies. We also do not rule out the possibility that these galaxies might have lost their gas through pre-processing rather than ram pressure stripping. 
\end{enumerate}

We utilized the availability of multi-wavelength data for galaxies in the Virgo cluster to study the star formation properties tracing different timescales. However, the parameters we extracted were based on integrated photometry, which limited a detailed study of the effects of ram pressure stripping on the ISM of the galaxies. To study the comprehensive picture of ram pressure stripping on the star formation activity of cluster galaxies, it is needed to make use of high spatial resolution data obtainable via integral field spectroscopy (IFS) observations. IFS data allows spatially resolved mapping of emission lines from which one can obtain diagnostic line ratios to constrain the ionization mechanism of the ISM, i.e., distinguish whether the observed emission originates from AGN feedback or star formation (e.g., \citealt{merluzzi13}). In particular, simultaneous observations of H$\alpha$ and H$\beta$ emission lines enable the derivation of dust extinction maps that can be studied along with H$\alpha$-derived SFR maps to study the star formation activity of galaxies on local scales \citep{nelson16, jafariyazani19}. 


\acknowledgments
We thank the referees for their insightful comments that helped improve the paper. This study was supported by the National Research Foundation of Korea (NRF)  grant  funded  by  the  Korean  Government  (NRF-2019R1A2C2084019). Funding for the SDSS and SDSS-II has been provided by the Alfred P. Sloan Foundation, the Participating Institutions, the National Science Foundation, the U.S. Department of Energy, the National Aeronautics and Space Administration, the Japanese Monbukagakusho, the Max Planck Society, and the Higher Education Funding Council for England. The SDSS Web Site is http://www.sdss.org/. The SDSS is managed by the Astrophysical Research Consortium for the Participating Institutions. The Participating Institutions are the American Museum of Natural History, Astrophysical Institute Potsdam, University of Basel, University of Cambridge, Case Western Reserve University, University of Chicago, Drexel University, Fermilab, the Institute for Advanced Study, the Japan Participation Group, Johns Hopkins University, the Joint Institute for Nuclear Astrophysics, the Kavli Institute for Particle Astrophysics and Cosmology, the Korean Scientist Group, the Chinese Academy of Sciences (LAMOST), Los Alamos National Laboratory, the Max-Planck-Institute for Astronomy (MPIA), the Max-Planck-Institute for Astrophysics (MPA), New Mexico State University, Ohio State University, University of Pittsburgh, University of Portsmouth, Princeton University, the United States Naval Observatory, and the University of Washington. This publication makes use of data products from the Wide-field Infrared Survey Explorer, which is a joint project of the University of California, Los Angeles, and the Jet Propulsion Laboratory/California Institute of Technology, and NEOWISE, which is a project of the Jet Propulsion Laboratory/California Institute of Technology. WISE and NEOWISE are funded by the National Aeronautics and Space Administration.



\end{document}